\documentclass[manuscript]{acmart} 

\usepackage{soul}
\usepackage{float} 
\usepackage{enumitem}

\usepackage[symbol]{footmisc}

\usepackage{wrapfig}
\usepackage{colortbl}
\usepackage[dvipsnames]{xcolor} 
\usepackage{placeins} 
\usepackage{amsmath} 
\usepackage{algorithm}
\usepackage{algpseudocode} 
\usepackage{multirow}
\usepackage{subcaption}
\usepackage{booktabs}

\usepackage{booktabs}
\usepackage{hyperref} 
\hypersetup{
    colorlinks=true,
    linkcolor=black,
    citecolor=CadetBlue,
    filecolor=CadetBlue,      
    urlcolor=CadetBlue,
}
\usepackage{multicol} 

\usepackage{color,soul}
\sethlcolor{yellow}

\newcommand{\cem}{CE\textsuperscript{-}}
\newcommand{\ccem}{CCE\textsuperscript{-}}
\newcommand{\ce}{CE}
\newcommand{\cce}{CCE} 
\newcommand{\bce}{BCE}
\newcommand{\sce}{SCE}
\newcommand{\sr}{SRS}
\newcommand{\vs}{|V|}
\newcommand{\vecT}{\mathbf{t}}
\newcommand{\vecP}{\mathbf{p}}

\begin{document}


\title{Faster and Memory-Efficient Training of Sequential Recommendation Models for Large Catalogs}

\author{Maxim Zhelnin}
\authornote{Both authors contributed equally to this work.}
\email{zhelninmax@gmail.com}
\affiliation{%
  \institution{MWS AI}
  \city{Moscow}
  \country{Russia}}

\author{Dmitry Redko}
\authornotemark[1]
\email{dmitryredko444@gmail.com}
\affiliation{%
  \institution{Applied AI}
  \city{Moscow}
  \country{Russia}}

\author{Daniil Volkov}
\email{danvol121@gmail.com}
\affiliation{%
  \institution{Applied AI}
  \city{Moscow}
  \country{Russia}}

\author{Anna Volodkevich}
\email{volodkanna@yandex.ru}
\affiliation{%
  \institution{Sber AI Lab, Applied AI}
  \city{Moscow}
  \country{Russia}}

\author{Petr Sokerin}
\authornote{Corresponding author.}
\email{sokerinpo@gmail.com}
\affiliation{%
  \institution{Applied AI}
  \city{Moscow}
  \country{Russia}}

\author{Valeriy Shevchenko}
\email{escape756@gmail.com}
\affiliation{%
  \institution{IVI}
  \city{Moscow}
  \country{Russia}}

\author{Egor Shvetsov}
\authornotemark[2] 
\email{e.shvetsov@applied-ai.ru}
\affiliation{%
  \institution{Applied AI}
  \city{Moscow}
  \country{Russia}}

\author{Alexey Vasilev}
\email{alexxl.vasilev@yandex.ru}
\affiliation{%
  \institution{Sber AI Lab}
  \city{Moscow}
  \country{Russia}}

\author{Darya Denisova}
\email{duny.explorer@gmail.com}
\affiliation{%
  \institution{Sber AI Lab}
  \city{Moscow}
  \country{Russia}}

\author{Ruslan Izmailov}
\email{lord.rik@yandex.ru}
\affiliation{%
  \institution{Sber}
  \city{Moscow}
  \country{Russia}}

\author{Alexey Zaytsev}
\email{likzet@gmail.com}
\affiliation{%
  \institution{Applied AI}
  \city{Moscow}
  \country{Russia}}

\renewcommand{\shortauthors}{Zhelnin et al.}





\maketitle

\textbf{Abstract.} 
Sequential recommendations (SR) with transformer-based architectures are widely adopted, and SR models require frequent retraining to adapt to changing user preferences. However, training transformer-based SR models incurs high computational cost from scoring large item catalogs (often thousands of items). This is mainly due to cross-entropy loss, where peak memory scales with catalog size, batch size, and sequence length. To reduce memory, practitioners combine cross-entropy (CE) with negative sampling, lowering explicit memory needs of the final layer. Yet few negative samples degrade performance; increasing negatives and batch size improves results but quickly exceeds industrial GPU memory (~40Gb).

In this work, we introduce \ccem{}, a GPU-efficient implementation of CE with negative sampling. \ccem{} accelerates training by up to 2× and reduces memory by more than 10×. The saved memory enables higher accuracy on datasets with large catalogs versus models trained with standard PyTorch losses. We also analyze key memory-related hyperparameters and show the need to balance them: scaling both the number of negative samples and batch size yields better results than maximizing only one. To support adoption, we release a Triton kernel implementing the method efficiently.\footnote{Code, reproducibility materials, and all scripts for generating figures are available at 
\url{https://github.com/On-Point-RND/MemoryEfficientSRS}}

\section{Introduction}
\label{sec:introduction}

User actions can be modeled sequentially, conditioning on past actions in diverse scenarios, ranging from a mechanical series of actions in a banking app \cite{boka2024survey} to a mood- and neighborhood-influenced selection of music tracks \cite{krause2014music}.
Consequently, Sequential Recommender Systems (\sr{}), which model user interactions as sequences, have become essential components of modern recommendation pipelines. \sr{} effectively incorporates ideas from natural language processing (NLP) through attention-based architectures such as SASRec~\cite{kang2018self} and BERT4Rec~\cite{sun2019bert4rec}, as well as the cross-entropy loss function~\citep{klenitskiy2024does, sun2019bert4rec, mezentsev2024scalable}.


The deployment of \sr{} often faces two challenges: (1) high memory demands due to large catalogs, (2) a need for frequent retraining to acquire recent user actions. 
Although peak memory usage was extensively addressed in the literature, training speed did not receive as much attention. In this paper, we consider both problems. 


\paragraph{Peak memory} 
Peak memory is one of the main computational constraints when cross-entropy loss (\ce) is applied to large catalogs. The memory footprint at the final layer for logits can be estimated as \( bs \cdot sl \cdot |V| \)\footnote[1]{To illustrate, consider a batch of 256 sequences, each of length 100, with a catalog containing \(10^6\) items. The resulting logit tensor would require the storage of \(256 \cdot 100 \cdot 10^6\) elements, equating to approximately \emph{100 GB of memory}.}, where $bs$ is the batch size, $sl$ sequence length and \(|V|\) is the catalog or vocabulary size, typically the most significant contributor. This final layer uses up to 90\% of the combined memory footprint, gradients, and activations, when training modern Large Language Models (LLMs)~\cite{wijmans2024cut}. However, in \sr{}, \(|V|\) can reach millions of items~\cite{zivic2024scaling}, surpassing even LLMs, which have vocabulary sizes orders of magnitude smaller: 32K, 152K, and 256K for Llama2-7B~\cite{touvron2023llama}, QWEN~\cite{bai2023qwen}, and Gemma-7B~\cite{team2403gemma}, respectively.

This memory footprint has been a key driver for early \sr{} adopting the binary cross-entropy loss (\bce{})~\cite{kang2018self} and the cross-entropy loss with negative sampling (\cem{})~\cite{sun2019bert4rec, klenitskiy2023turning}. \cem{} directly addresses memory constraints caused by large item catalogs~\cite{klenitskiy2023turning, mezentsev2024scalable}. Instead of scoring all items during training, a subset of negative samples (\(ns\)) is used, where \(ns \ll |V|\), to approximate the full \ce{} loss. This reduces computational complexity and memory requirements but introduces a hyperparameter trade-off between batch size, sequence length, and number of negative samples (\textit{ns}) in memory-constrained scenarios. Prior work indicates that maximizing $ns$ in \cem{} or using \ce{} can improve performance of the models~\cite{di2024theoretical, klenitskiy2023turning, guo2024scaling}. However, empirical results demonstrate that \cem{} can sometimes outperform full \ce{} for some datasets~\cite{mezentsev2024scalable, klenitskiy2023turning} and the performance of models trained with \cem{} shows a non-linear dependence on \(ns\)\cite{prakash2024evaluating}. This creates an open question of whether practitioners should maximize $ns$ given a fixed memory budget or prioritize other memory-related parameters. We will refer to this question as the first question \textbf{Q1}.

\paragraph{Training Speed}
Although sampling strategies such as \cem{} reduce memory consumption, they also improve training speed by requiring fewer operations, partially addressing the problem of frequent retraining \sr{}.  However, there are other approaches for faster training: quantization~\cite{wortsman2023stable},  activation sparsification~\cite{hu2024accelerating}, and finally gradient sparsification~\cite{alistarh2018convergence}. The latter one is usually overlooked, even though gradients usually require twice as many computations as the forward path. In our work, we demonstrate that gradient sparsification is especially suitable for \sr{}, provide a formal analysis in \S\ref{sec:sparse_grads}, and study how sparse gradients accelerate model training and their impact on accuracy degradation. We will refer to it as our second research question \textbf{Q2}.

\begin{figure}[!h]
  \centering
  \includegraphics[width=0.98\textwidth]{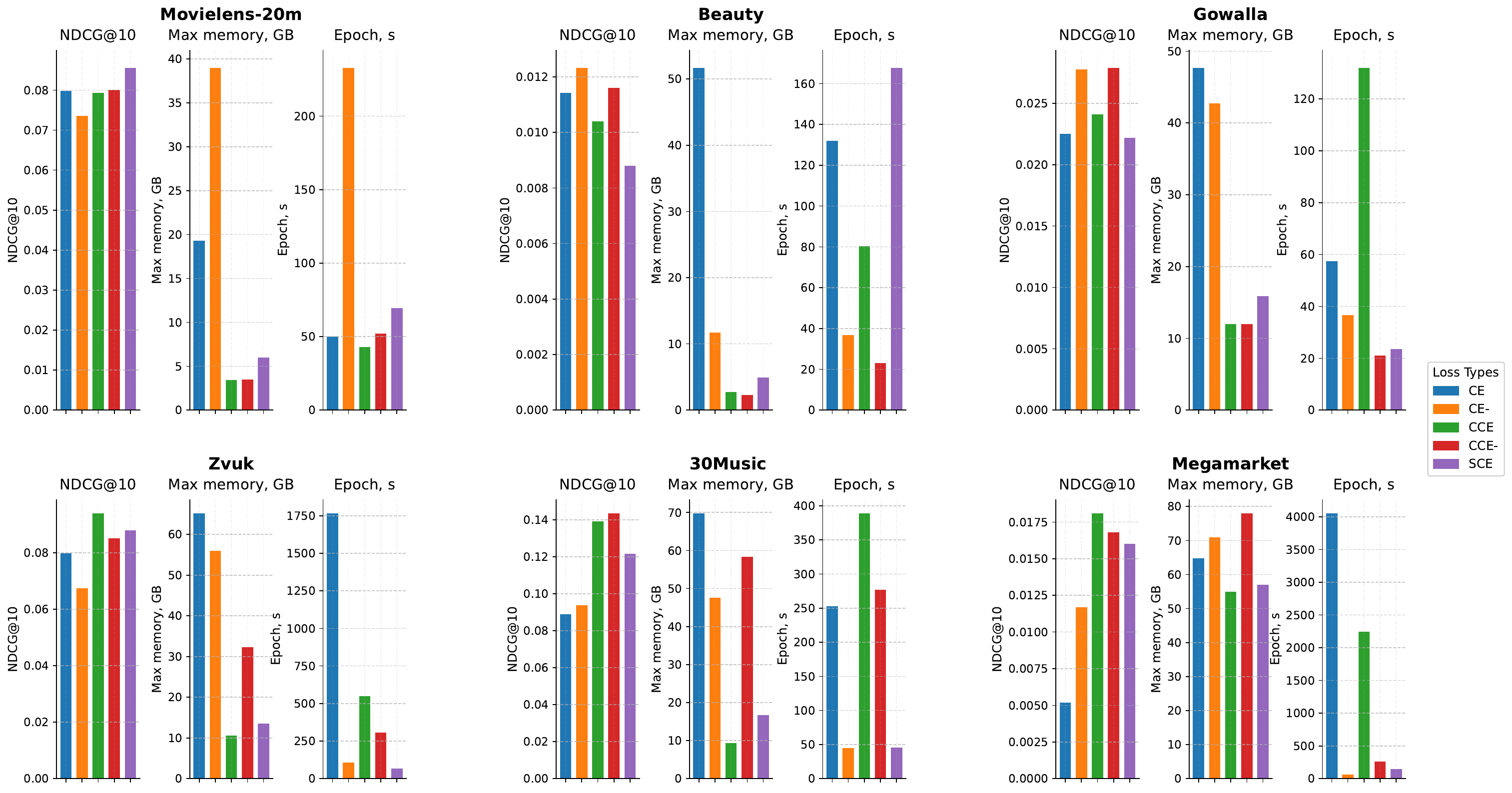}
  \caption{Here, we compare several methods \ce{}, \cem{}, \cce{}, \ccem{}, and \sce{} in terms of  (1) \textbf{NDCG@10}, (2) \textbf{Training time} per epoch in seconds, and (3) \textbf{Memory consumption} in Gb for the SASRec model across six datasets. We demonstrate the best performance achieved through optimized hyperparameters. We can see that \ce{}, \cem{} require much more time and memory usage. \textbf{Acceleration and memory consumption is measured against float 16 (mixed-precision) baseline.}}
  \label{fig:best_metrics}
  \Description{description}
\end{figure}

\paragraph{Hardware-friendly Low-Level Efficient Implementations}
Computational efficiency critically depends on the co-design of architecture, hardware, and low-level implementations. While hardware is often fixed, efficiency can be optimized through architectures better aligned with underlying hardware~\cite{shvetsov2024quantnas} or specialized low-level optimizations. Though highly specific, such optimizations yield significant computational benefits. Examples include FlashAttention~\cite{dao2022flashattention} for efficient attention computation and Liger kernels~\cite{hsu2024liger} for optimized \ce{} computation. Among them,  Cut-Cross Entropy (\cce{}) was recently developed for LLMs; it helps to alleviate memory bottlenecks during training when \ce loss is employed. However, until now, the \cce{}  has remained unexplored for \sr{} models. Furthermore, our analysis reveals the importance of negative sampling, which is not supported by the original \cce{}. Development and analysis of \cce{} with negative sampling is our third research question - \textbf{Q3}, more precisely: Can we design a \cce{} alternative that is faster and more memory efficient for \sr{}?





Our work addresses these questions through the following contributions:

\begin{enumerate}[leftmargin=0.5cm]
    \item \textbf{Comprehensive Memory-Accuracy Trade-off Analysis}:  
    To address \textbf{Q1}, we conducted a systematic investigation into the relationships among accuracy of SASRec, batch size (\(bs\)), sequence length (\(sl\)), the number of negative sample size (\(ns\)) across six datasets with varying item catalog sizes (\(|V|\)). Our findings indicate that simply maximizing \(ns\) in \cem{} or employing \ce{} to encompass the entire item catalog \(|V|\) leads to suboptimal performance in memory-constrained environments. Instead, the choice between \cem{} and \ce{} requires a data set-specific calibration. We provide analysis for configuring these parameters to balance accuracy and memory efficiency.  
    
    Furthermore, we make publicly available a dataset comprising 997 data points, each representing SASRec trained across the considered datasets. For every configuration, we log \(bs\), \(sl\), \(ns\), memory consumption, and various metrics along the training steps. This dataset enables reproducible analysis of memory-accuracy trade-offs.

    \item \textbf{Sparse Gradients For training acceleration}:  
    To address \textbf{Q2}, we conducted an analysis of the classification layer and observed that the matrices involved in gradient computations contain numerous elements with low magnitude. Due to data format limitations\footnote{The smallest value representable in FP16 is \(5.96 \cdot 10^{-8}\).}, this leads to \textbf{high sparsity} of the gradient matrices. Our experiments demonstrate that further gradient filtering preserves model quality while accelerating training. The mathematical analysis in \S\ref{sec:sparse_grads} further confirms that this phenomenon should generalize across sequential recommendation (\sr{}) models trained with cross-entropy (\ce{}) loss and large catalog sizes.

    \item \textbf{Development of Faster \ccem{} kernel}: We are the first to adapt and demonstrate that \cce{} is exceptionally well-suited for training \sr{} models when the \ce{} loss is required. Furthermore, motivated by the findings of our first contribution on the significance of negative sampling for certain data sets, we developed \ccem{}, a GPU-optimized variant of the loss \ce{} with negative sampling. We demonstrate that \ccem{} outperforms \cce{} on 4/6 datasets and achieves up to 89\% faster training as it is presented in Figure \ref{fig:best_metrics}. \textbf{Acceleration and memory consumption is measured against float 16 (mixed-precision) baseline.}


    \item \textbf{Memory-Efficient Metric Improvement}: 
    By reallocating the memory savings achieved through the use of \cce{} and \ccem{} to increase batch size  \(bs\),  sequence length \(sl\), and the number of negative samples \(ns\), we attain up to a 30\% improvement in accuracy while maintaining the same computational budget.

\end{enumerate}

\textbf{Paper Organization:}
\S\ref{sec:related_work} reviews related work on loss functions and 
efficient training for sequential recommendation.
\S\ref{sec:cce_methods} introduces CCE and CCE\textsuperscript{-}, 
describes the sparse gradient phenomenon, and provides a 
formal analysis.
\S\ref{sec:experiments} describes the experimental setup, datasets, 
and evaluation methodology.
\S\ref{sec:scaling} presents the main results: the memory--accuracy 
trade-off analysis (Q1), gradient filtering experiments (Q2), 
and a comprehensive comparison of CCE and CCE\textsuperscript{-} 
against CE, CE\textsuperscript{-}, BCE, and SCE baselines. Appendix~\ref{app:alternatives} compares other approaches such as 
quantization and sparsification.
Appendix~\ref{app:additional_results} provides additional figures that were not included in the main article, as well as the results for BERT4Rec and the estimated optimal $bs$/ $ns$ ratios per dataset.

\begin{table*}[!t]
  \centering
  \small
   \caption{NDCG@10 metric values for the SASRec model trained using different loss functions. The best metric value is in bold, and the second-best is underlined.}
  \label{tab:total_metrics}
  \begin{minipage}{\textwidth}
    \centering
    \begin{tabular}{lccccc}
        \hline
      Dataset & BCE & CE & CE- & CCE & \textbf{CCE-} \\
      \hline
      Megamarket & 0.0017 & 0.0052 & 0.0117 & \textbf{0.0181} & \underline{0.0168} \\
      Zvuk       & 0.0046 & 0.0797 & 0.0675 & \textbf{0.0940} & \underline{0.0852} \\
      30Music    & 0.0004 & 0.0891 & 0.0939 & \underline{0.1393} & \textbf{0.1436} \\
      Gowalla    & 0.0197 & 0.0225 & \underline{0.0278} & 0.0241 & \textbf{0.0279} \\
      Beauty     & 0.0094 & 0.0114 & \textbf{0.0123} & 0.0104 & \underline{0.0116} \\
      Movielens-20m     & \textbf{0.0557} & 0.0510 & 0.0513 & 0.0511 & \underline{0.0523} \\
 \hline
    \end{tabular}
  \end{minipage}
\end{table*}

\section{Related work}
\label{sec:related_work}
 \textbf{Cross-Entropy and sampled \ce{} for \sr{}.}  
    Sequential recommendation has seen significant advancements in recent years. One pivotal observation from ~\cite{klenitskiy2023turning} demonstrated that SASRec's performance improves when trained with cross-entropy loss instead of binary cross-entropy. Unfortunately, computing cross-entropy over all items is computationally prohibitive. To address this, researchers adopted \textit{sampled cross-entropy loss}, which uses a subset of negative items. This approach was theoretically justified in ~\cite{di2024theoretical}, showing that increasing negative samples enhances performance. \textbf{However, negative sampling has non-linear performance scaling.}  
    Despite its benefits, negative sampling amplifies popularity bias, as popular items are more likely to be sampled ~\cite{prakash2024evaluating}. Consequently, model performance becomes highly dependent on dataset distribution and sampling strategy. Furthermore, ~\cite{prakash2024evaluating} highlights that performance does not scale linearly with the number of negative samples, which complicates the selection of hyperparameters. To address this problem, we perform a careful and extensive study of memory-dependent \textit{hyperparameter analysis for optimal accuracy-memory tradeoff} (\S\ref{sec:scaling}). 
    \textbf{Prior work on scaling} for \sr{} models ~\cite{guo2024scaling, zhang2024scaling, zivic2024scaling, xu2025efficient} emphasizes the benefits of scaling sequence length and model depth. However, our experiments reveal that \textbf{expanding a single factor}\footnote{Factors include batch size, sequence length, and number of negative samples.} \textbf{yields limited gains}, underscoring the need for multi-dimensional scaling. Unlike ~\cite{mezentsev2024scalable}, our work focuses only on feature-scaling\footnote{Feature scaling refers to adjusting batch size, sequence length, etc.} effects while keeping model architecture fixed.

  \textbf{Computationally efficient approaches.}  
    Advanced sampling techniques for balancing accuracy and memory were proposed in ~\cite{mezentsev2024scalable}. Liger-Kernel ~\cite{hsu2024liger}, implemented in the Triton language ~\cite{tillet2019triton}, addresses memory consumption issues through the use of FusedLinearCrossEntropy (FLCE). While theoretical convergence guarantees for sparse gradient methods were established in ~\cite{alistarh2018convergence}, it's application for efficient training remains limited ~\citep{chekalina2024sparsegrad, wijmans2024cut}. Building on FLCE memory-efficient cross-entropy with saturated gradient filtering beyond numerical precision was proposed (\cce{}) for LLMs ~\cite{wijmans2024cut}. Since saturated gradients do not fit numerical precision and turn to zero, their removal does not affect model convergence. In this work, we analyze whether the removal of low-magnitude non-zero gradients would help to accelerate model training with minimal performance degradation.

  \textbf{Possible further directions and trends.}  
  Even when isolating experiments to feature scaling, there remains significant flexibility in training strategies. For example, work ~\cite{li2022stability} demonstrated that training GPT-like models with short sequences initially and gradually increasing them enables stable, efficient training with larger batch sizes and learning rates. Similarly, authors of ~\cite{di2024theoretical} found that starting with fewer negative samples and increasing their count later boosts \sr{} model performance. Moreover, based on our results, an adaptive threshold to prune gradients can be employed.    While these adaptive strategies are promising, we focus on studying simpler non-adaptive approaches to establish a stronger foundation for future research. 

\section{Methods}

\begin{figure}
    \centering
    \includegraphics[width=1\linewidth]{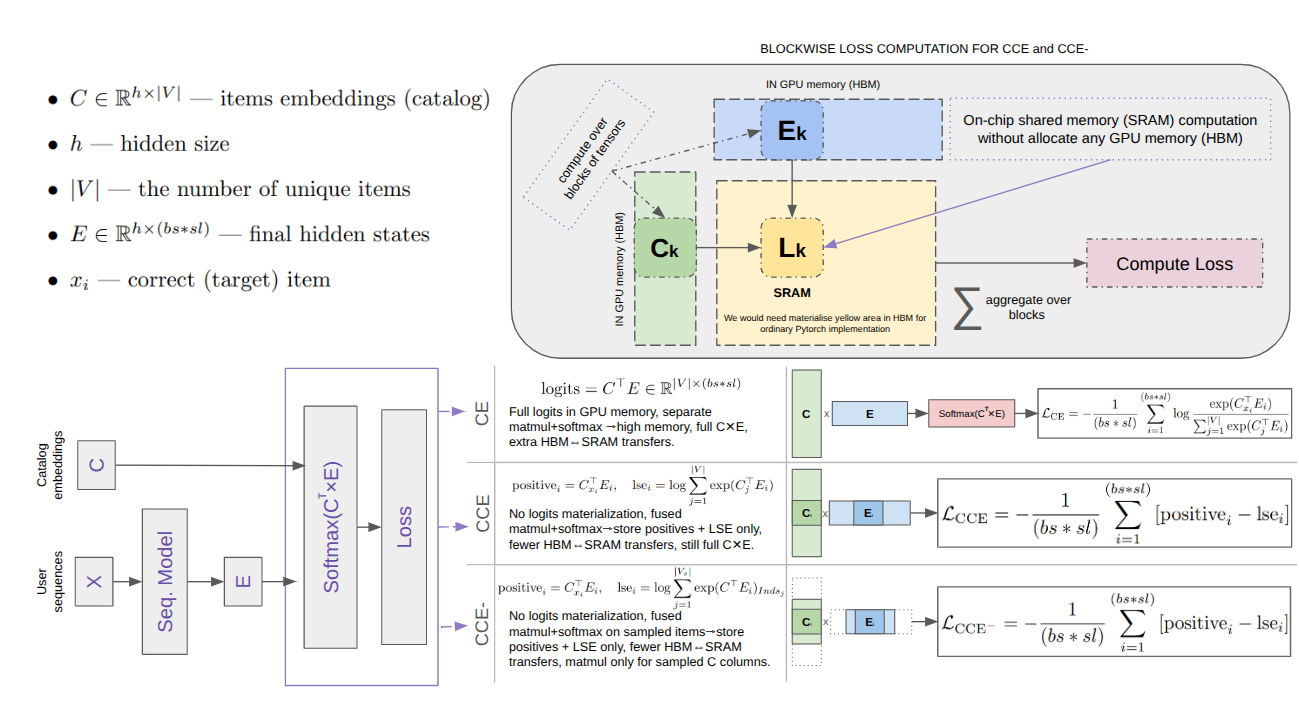}
    \caption{Illustration of \ce{}, \cce{} and \ccem{} implementation. \textbf{\ce{}}: Materializes full logits in GPU memory, matmul and softmax are computed separately $\rightarrow$ high GPU memory usage, full $C \times E$ matmul, extra data transfers between HBM and SRAM lead to higher time delay.  \textbf{\cce{}} : No logits materialization, fused matmul + softmax  $\rightarrow$ store only positive logits and LSE vector, reduced HBM $\leftarrow \rightarrow$ SRAM transfers, but still full  CxE matmul. \textbf{\ccem{}}: No logits materialization, fused matmul + softmax on sampled items $\rightarrow$ store only positive logits and LSE vector, reduced HBM $\leftarrow \rightarrow$ SRAM transfers, matmul only for sampled columns of C and corresponding E. }
    \label{fig:placeholder}
    \Description{description}
\end{figure}

\subsection{Cut-Cross-Entropy for RecSys}
\label{sec:cce_methods}


Our proposed method, \ccem{}, builds upon \cce{}, which itself extends the Liger-Kernel (LG)~\cite{hsu2024liger}.
Below, we provide a brief review of both \ccem{} and LG, highlighting their key differences.
We then focus on how \ccem{} introduces specific improvements tailored for sequential recommendation systems.


During training, we deal with two types of memory: GPU memory (High Bandwidth Memory, HBM) and on-chip shared memory (Static Random-Access Memory, SRAM). 
HBM is large but relatively slow and is used to store data for computations. 
GPU kernel loads input data from HBM into SRAM to perform computations. 
SRAM is significantly faster than HBM but has much less capacity. 
Therefore, only small chunks of input data can be loaded into SRAM at a time. Once computations are complete, the output is saved back to HBM. 
While there are more details, see~\cite{dao2022flashattention}, ~\cite{wijmans2024cut}, this structure shapes the main goal: reduce the amount of data loaded into GPU memory.
It can be achieved by minimizing the overall amount of memory used by a model or optimizing transfers between HBM and SRAM.

The main idea behind LG is that it processes hidden states in smaller chunks, therefore reducing peak memory usage.
It applies a linear classification head to each chunk, computes the logits, and then feeds these logits into a GPU to calculate partial losses with a special kernel. 
These losses are used to estimate the gradients for the weights of the classification head and the chunked hidden states.
While chunking reduces memory usage, increasing the number of chunks introduces latency overhead.

\cce{} enhances LG through the fusion of logit and softmax computations within a single GPU kernel. 
By executing intermediate operations in the high-speed on-chip shared memory (SRAM), this approach only requires the materialization of logits for correct items and the log-sum-exp (LSE) vector. 
This eliminates the need to store the entire logit tensor in the global GPU memory, resulting in a substantial reduction in GPU memory usage during both forward and backward propagation. To implement this optimization, new Triton kernels were developed, effectively balancing computational performance and memory efficiency.


In this study, we incorporated \cce{} into sequential recommendation transformer models to evaluate its effectiveness for recommendation tasks. For this purpose, the three-dimensional output tensor of embeddings produced by the final transformer block is flattened into a matrix $\mathbf{E} \in \mathbb{R}^{N\times D}$, where $D$ represents the hidden size and $N$ corresponds to the total number of items, calculated as the product of the batch size $bs$ and the sequence length $sl$. The CE loss for the set of items of size $\vert V \vert$ is written as follows: 
\begin{align}
    \mathcal{L} &= -\dfrac{1}{N}\sum_{i=1}^{N}\mathbf{l}_{i}, \\
    \mathbf{l} &= (\mathbf{C}^{\mathrm{T}} \mathbf{E})_{\mathbf{x}} - \mathrm{log} \sum_{j=1}^{ V} \exp{(\mathbf{C}^{\mathrm{T}} \mathbf{E})},
\label{eqn: cce_section_CE_vec}
\end{align}
where $\mathbf{C} \in \mathbb{R}^{D \times \vert V \vert }$ represents the linear classifier, projecting the embeddings $\mathbf{E}$ into the output space. 
The first term $(\mathbf{C}^{\mathrm{T}}\mathbf{E})_{\mathbf{x}}$ corresponds to the vector of logits for the correct items with indices $\mathbf{x}=(x_{1}, \ldots, x_{N})$, which is computed as 
$(\mathbf{C}^{\mathrm{T}}\mathbf{E})_{\mathbf{x}} = [C_{x_{1}}^{\mathrm{T}}E_{1}, \ldots, C_{x_{N}}^{\mathrm{T}}E_{N}]$, $C_{x_{j}}$ refers to the columns of matrix $\mathbf{C}$ associated with the correct items $x_{j}$, and $E_{j}$ denotes columns of the output embeddings matrix. The second term $\mathrm{log}\sum_{j=1}^{V}\exp{(\mathbf{C}^{\mathrm{T}}\mathbf{E})}$ represents a combination of the log-sum-exp operation and matrix multiplication.

Formulating the \ce{} loss as Equation \ref{eqn: cce_section_CE_vec} allows it to be decomposed into calculating logits for correct items and performing a log-sum-exp operation on the full logits matrix $\mathbf{C}^{T}\mathbf{E}$. Consequently, all operations in the second term can be fused into a single GPU kernel, thereby avoiding the materialization of the large full logits matrix in the global GPU memory.

Besides, to accelerate computations during the backward pass, the Triton kernels for \cce{} are capable of performing gradient filtering at the final layer. 
This feature facilitates the efficient handling of sparsity in gradient matrices, which are common in recommender systems, as discussed in  \S\ref{sec:sparse_grads}. Filtering is achieved by disregarding computations with values below a prescribed threshold. The default threshold is set to the numerical precision of the data format, although other values can also be used. Consequently, \cce{} not only reduces the memory footprint, but also accelerates training due to ignoring computations involving near-zero values.




\begin{figure}
  \includegraphics[width=0.5\linewidth]{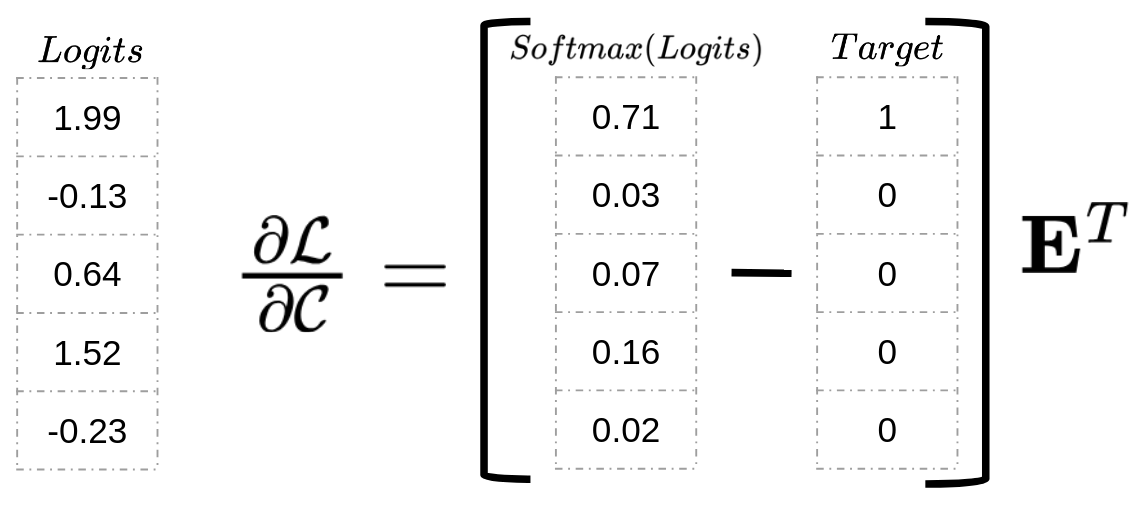}
  \caption{An illustration of computing gradients for the weights of the final layer.}
  \label{fig:grad_computation}
  \Description{description}
\end{figure}

\subsection{Sparse Gradients in Training with \ce{} loss}
\label{sec:sparse_grads}

In the recommender systems, the objective is to predict the next item from a predefined set $V$ of \( \vs{} \) items. Similar to the next-token prediction, there is exactly one correct class and \( \vs{} - 1 \) incorrect classes. Thus, employing loss functions such as the \ce{} loss provides a suitable estimate of model quality ~\citep{di2024theoretical}.

Let us consider the gradient sparsity that arises in the final layer of a sequential recommendation model during training with the \ce{} loss.

Let $\mathcal{L} = \mathcal{L}(\vecP, \vecT)$ denote the \ce{} loss, that is used to measure discrepancy between model predictions $\vecP$ and target one-hot encoded labels $\vecT$. The predicted probability vector  $ \vecP$ is computed as:
$$
\mathbf{p} = \frac{\exp(\mathbf{C}^T E)}{\sum_{j=1}^{|V|} \exp(C^{T}_{j} E)},
$$
where $\mathbf{C}$ is the weight matrix of the final layer, $C_{j}$ refers to the columns of the matrix $\mathbf{C}$, and $E$ is the embedding vector.

For the incorrect label \(y\), the gradient is $\frac{\partial \mathcal{L}}{\partial C_{yj}}  = p_ye_j$, where $e_{j}$ is a component of the vector $E$, see Figure~\ref{fig:grad_computation}. When the item catalog size $|V|$ is large, the probability of incorrect labeling $p_{y}\rightarrow0$ since the softmax function was applied, and so $\frac{\partial \mathcal{L}}{\partial C_{yj}} \rightarrow 0 $. Because we have $|V|-1$ incorrect labels, most gradients remain near zero or saturate to zero based on data format, for example, for FP16 precision, the smallest positive number is \(6 \times 10^{-8}\).

\begin{figure}[ht]
  \centering
  \includegraphics[width=0.75\textwidth]{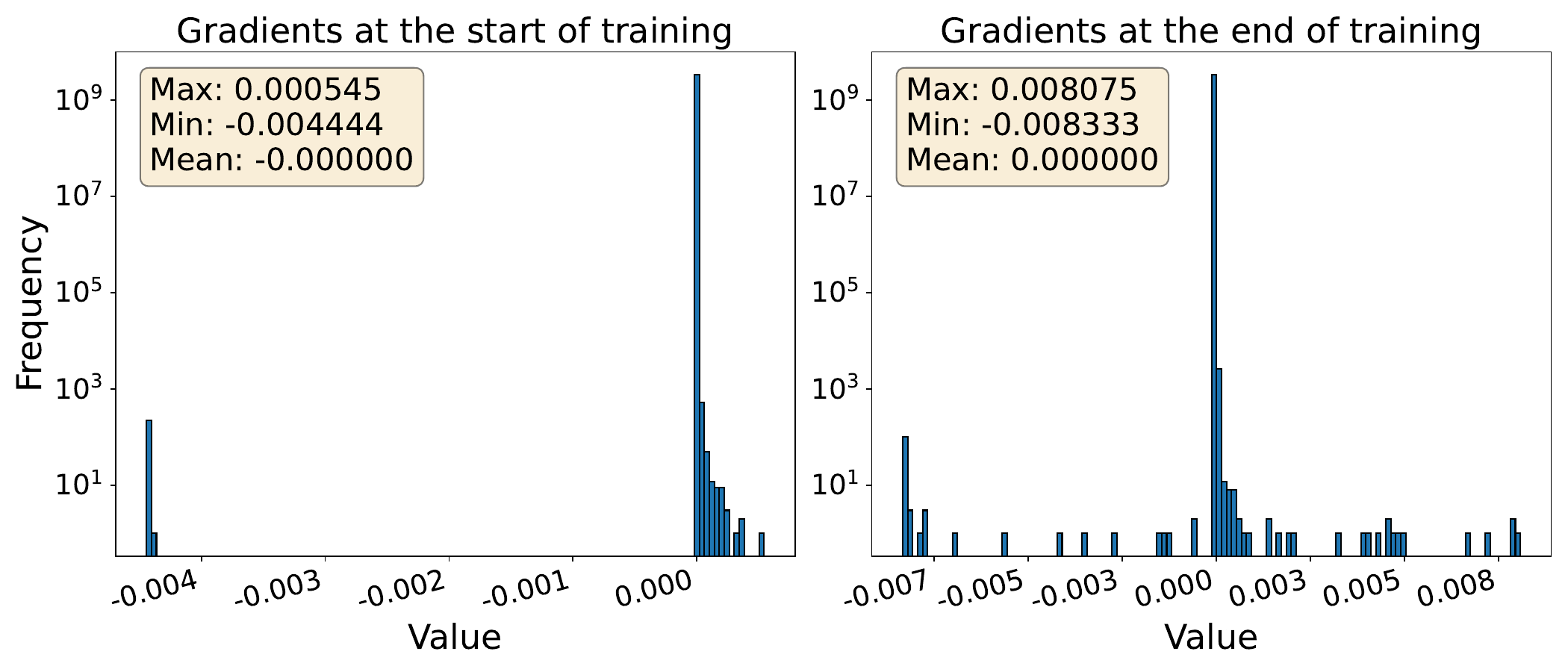}
\caption{Histograms of the gradient distributions at the final layer at the beginning (left) and end (right) of the training, the Megamarket dataset.}
\label{fig:sparse_grads_training}
  \Description{description}
\end{figure}

\textbf{Empirical Evaluation.} To validate our analysis, we present histograms of gradient distributions before and after training on the Megamarket dataset. As shown in Figure~\ref{fig:sparse_grads_training}, gradient values cluster near zero in both phases, confirming the persistence of sparsity throughout training.

\subsection{Cut-Cross-Entropy with Negative Sampling}
As demonstrated in \citep{klenitskiy2023turning}, negative sampling offers dual advantages: it allows better accuracy on certain datasets \textit{and} accelerates training compared to full vocabulary utilization. Motivated by this, we extend \ccem{} by integrating negative sampling into it.

\textbf{\ccem{} Loss Formulation.}  The loss function operates on an embedding matrix $\mathbf{E} \in \mathbb{R}^{N \times D}$, which is obtained by flattening the 3D output tensor along its first two dimensions. During training, at each time step, the positive items corresponding to users are augmented with a dynamically sampled set of $ns$ negative items. This creates a compact item subset $V_s \subset V$ where:  $|V_s| = 1 + ns$ (1 positive + $ns$ negatives) and  $|V_s| \ll |V|$ (sampled subset size $\ll$ size of the full catalog of items ).
The loss calculation is thereby confined to interactions within this compact subset, enabling efficient training, as discussed below.
The \ccem loss is written as

\begin{equation}
    \mathcal{L}_{NS} = -\dfrac{1}{N}\sum_{i=1}^{N}\mathbf{l}^{NS}_{i},
\end{equation}

\begin{equation}
    \mathbf{l}^{NS} = (\mathbf{C}^{\mathrm{T}}\mathbf{E})_{\mathbf{x}} - \mathrm{log}\sum_{j=1}^{V_{s}}\exp{(\mathbf{C}^{\mathrm{T}}\mathbf{E})_{\mathbf{Inds}_{j}}}.
\label{eqn: cce_section_CE_negative_loss_vec}
\end{equation}

The first term is identical to that in Eq. \ref{eqn: cce_section_CE_vec}. The second term is a vector defined as 
\begin{equation}
   LSE =\mathrm{log}\sum_{j=1}^{V_{s}}\exp{(\mathbf{C}^{\mathrm{T}}\mathbf{E})_{\mathbf{Inds}_{j}}}, 
\label{eqn: cce_section_indexed_LSE}
\end{equation}
which incorporates the log-sum-exp function and indexed matrix multiplication. This is expressed as $(\mathbf{C}^{\mathrm{T}}\mathbf{E})_{Inds_{j}} = [C_{x_{1j}}^{\mathrm{T}} E_{1}, \ldots, \mathbf{C}_{x_{Nj}} E_{N}]$, where the set $Inds_{j} = (x_{1j}, \ldots, x_{Nj})$ represents the indices of the columns of the matrix $\mathbf{C}$ used for negative sampling. Specifically, when $j=1$, the columns correspond to the correct items, while for $1 < j < |V_{s}|+1$, the columns are related to negative examples associated with the embeddings $\mathbf{E} = [E_{1}, \ldots, E_{N}]$ of the user items.

The correct logits are represented by the first term of Eq. \ref{eqn: cce_section_CE_negative_loss_vec} are materialized in global GPU memory and computed using an identical \cce{}  kernel. The second term in the equation involves several operations: indexing $\mathbf{C}$ based on $\mathbf{Inds}_{j}$, performing the dot product and summing along the inner dimension for $\mathbf{C}_{x_{ij}}^{\mathrm{T}} E_{i}$, and computing the log-sum-exp operation.

To achieve a memory-efficient computation for this term, we developed a specialized kernel that fuses these operations. For this purpose, we adopt an approach for the online softmax computation. As a result, all intermediate operations are carried out on-chip using shared memory, while only the log-sum-exp vector is stored in the global memory~\citep{milakov2018onlinesoftmax, hsuliger}.

During the backward pass, the gradients for the output embeddings and the matrices of the linear classifier are computed as follows:
\begin{equation}
    \nabla E_{i} = \sum_{j=1}^{|V_{s}|}(\mathbf{S}\cdot \nabla LSE)_{ij} C_{x_{ij}}, \quad
    \nabla C_{x_{ij}}^{\mathrm{T}} = (\mathbf{S}\cdot \nabla LSE)_{j}^{\mathrm{T}} \cdot E_{i},
\label{eqn: cce_section_CE_negative_loss_grad}
\end{equation}
where softmax $\mathbf{S} = \mathrm{softmax}(\mathbf{C}^{\mathrm{T}} \mathbf{E})_{\mathbf{Inds}}$ and $\cdot$ denotes the row-by-row element-wise multiplication. As in the \cce{}, the matrices $\mathbf{S}$ and $\mathbf{S}\cdot \nabla LSE$ are not materialized in global memory; instead, most computations are performed in shared memory. 
We recompute $(\mathbf{C}^{\mathrm{T}}\mathbf{E})_{\mathbf{Inds}}$ at the backward pass and calculate the softmax $\mathbf{S}$ as $\mathbf{S}=\mathrm{softmax}(\mathbf{C}^{\mathrm{T}}\mathbf{E})_{\mathbf{Inds}}=\exp((\mathbf{C}^{\mathrm{T}}\mathbf{E})_{\mathbf{Inds}} - LSE)$.
Only the gradient $ \nabla LSE$ is retained in the global memory.

Algorithms \ref{alg: cce_minus_forward_pass} and \ref{alg: cce_minus_backward_pass} outline the computational process for the LSE vector from Equation \ref{eqn: cce_section_indexed_LSE} during the forward pass, the gradients from Equation \ref{eqn: cce_section_CE_negative_loss_grad} during the backward pass, and also clarify the data access scheme.

During the forward pass, the output embedding $\mathbf{E}$ is divided into the blocks $E_{n, D}$ of size $N_{B} \times D$. 
Since the inner dimension $D$ in transformer models for RecSys is typically not large (in ~\citep{kang2018self}, $D<50$; in ~\citep{klenitskiy2023turning},~\citep{klenitskiy2024does}, $D=64$; in ~\cite{guo2024scaling}, $D < 400$ ), division is not performed along the $D$ dimension. 
The matrix of the linear classifier $\mathbf{C}$ is loaded inside a for-loop by columns, determined by the block of indices $\mathbf{Inds}_{n, j}$ of size $N_{B}$.

 During the backward pass, each kernel is executed for the blocks of the output embedding matrix $E_{n, D}$ of size $N_{B} \times D$ and the indices matrix $\mathbf{Inds}_{n, j}$ of size $N_{B}$. 
After the computations, the gradient $\nabla \mathbf{C}$ contains non-zero values only in columns corresponding to the indices provided by $\mathbf{Inds}$.  Since \ccem{} aims to achieve acceleration by restricting the catalog size through negative sampling, we do not implement gradient filtering for this method, unlike in \cce{}. However, gradient filtering can be integrated after Step 4 of Algorithm \ref{alg: cce_minus_backward_pass} in a similar manner to \cce{}.


\begin{algorithm}
\caption{Memory-efficient linear-log-sum-exp with negative sampling, forward pass}
\begin{algorithmic}[1]
\State $\text{LSE} = -\infty_N$ \Comment{vectors of size $N$ in main GPU memory}
\For{blocks $\mathbf{E}_{n, D}$}
    \State $\mathbf{m} =  -\infty_n$, $\mathbf{d} =  \mathbf{0}_{n}$ \Comment{vectors of size $n$ in on-chip SRAM}
    \While{$j < V_{s} $}
        \State $\mathbf{c}_{n} = \mathbf{C}_{\mathbf{Inds}_{n, j}, D}$ \Comment{Indexed load of $\mathbf{C}$ into on-chip SRAM}
        \State $\mathbf{o}_n = \sum_{d=1}^{D} \mathbf{E}_{n, d} \cdot \mathbf{C}_{\mathbf{x}_{n}, d}$ \Comment{Blockwise scalar product}
        \State $\mathbf{m}_{new} = \max(\mathbf{m}, \mathbf{o})$ \Comment{Update the maximum value}
        \State $\mathbf{d} = \mathbf{d} \cdot \exp(\mathbf{m}_{new} - \mathbf{m}) +\exp(\mathbf{o}_n - \mathbf{m}_{new})$
         \Comment{Update the sum}
        \State $\mathbf{m} = \mathbf{m}_{new}$ \Comment{Reassign the maximum value}
    \EndWhile
    \State $\text{LSE}_{n} = \mathbf{m} + \log \mathbf{d}$ \Comment{Compute LSE values in on chip SRAM}
\EndFor
\end{algorithmic}
\label{alg: cce_minus_forward_pass}
\end{algorithm}

\begin{algorithm}
\caption{Memory-efficient linear-log-sum-exp with negative sampling, backward pass }
\begin{algorithmic}[1]
\For{all blocks $\mathbf{E}_{n, D}$, $\mathbf{Inds}_{n, j}$}
    \State $\mathbf{c}_{n} = \mathbf{C}_{\mathbf{Inds}_{n, j}, D}$ \Comment{Indexed load of $\mathbf{C}$ into on-chip SRAM}
    \State $\mathbf{o}_n = \sum_{d=1}^{D} \mathbf{E}_{n, d} \cdot \mathbf{C}_{\mathbf{x}_{n}, d}$ \Comment{Blockwise scalar product}
    \State $\mathbf{S} = \exp(\mathbf{o}_n - LSE_{n})$
    \Comment{Compute the softmax}
    \State $\nabla E_{n, D} += (\mathbf{S}\cdot \nabla LSE)_{nj}\cdot \mathbf{c}_{n}$ \Comment{Update gradient for output embeddings}
    \State $\nabla \mathbf{C}_{\mathbf{Inds}_{n, j}, D}^\top += (\mathbf{S}\cdot \nabla LSE)_{j}^{\mathrm{T}} \cdot E_{i}$ \Comment{Indexed update gradient for the linear classifier}
\EndFor
\end{algorithmic}
\label{alg: cce_minus_backward_pass}
\end{algorithm}

\section{Experimental Settings}
\label{sec:experiments}

\subsection{Datasets}


\begin{table}[htbp]
\caption{Statistics for used datasets after preprocessing}
\label{tab:datasets}
\centering
\small 
\begin{tabular}{lcccc}
\hline
Dataset & \multicolumn{3}{c}{Number of\phantom{$aaaaa$}} & Mean sequence \\
        & Items & Users & Interactions & length \\
\hline
Movielens-20m       & $15$K   & $138$K & $12181$K   & $88.22$ \\
Beauty      & $176$K  & $674$K & $5028$K    & $7.46$ \\
Gowalla     & $309$K  & $85$K  & $4661$K    & $54.98$ \\
Zvuk        & $893$K  & $371$K & $243412$K  & $656.93$ \\
30Music     & $909$K  & $44$K  & $25639$K   & $577.71$ \\
Megamarket  & $1661$K & $350$K & $114846$K  & $327.88$ \\
\hline
\end{tabular}
\end{table}


We use six datasets: 30Music ~\cite{turrin201530music}, Zvuk ~\cite{shevchenko2024variability}, MovieLens-20M ~\cite{harper2015movielens}, MegaMarket~\cite{shevchenko2024variability}, Beauty ~\cite{mcauley2015image}, and Gowalla ~\cite{cheng2013you}. 
Our main focus is on datasets with a large item catalog: Zvuk, MegaMarket, Gowalla, and 30Music, which include more than 300K items. 
Moreover, Zvuk and MegaMarket are some of the largest open datasets in the recommendation domain ~\cite{shevchenko2024variability}. 
MovieLens and Beauty are added because of their popularity, used in more than 14 and 45 papers according to~\cite{klenitskiy2024does}. For Beauty, we employ its third version, which has the largest item catalog. The characteristics of all datasets after preprocessing are in Table \ref{tab:datasets}.

\subsection{Evaluation Methodology}

\paragraph{Dataset Splitting:}  \textbf{Global Temporal Cutoff}: To ensure alignment with real-world sequential prediction scenarios, we use a global temporal split with the last interaction as a target and "by user" validation scheme~\citep{gusak2025time}. We split interactions at the 90\% timestamp quantile. All interactions \textit{before} this cutoff form the training/validation pool, while interactions \textit{after} constitute the test set. \textbf{Validation Set}: From the training pool, we randomly select 5\% of users. For each selected user, their \textit{last interaction} is withheld for validation. \textbf{Test Set}: For users in the post-cutoff test set, only their \textit{last interaction} is retained for evaluation, while all previous ones are passed as model input.

\paragraph{Metrics:} Following prior work ~\cite{anelli2019discriminative, zhao2022revisiting, shevchenko2024variability}, we prioritize \textbf{NDCG@10} due to its strong correlation with overall ranking quality. Along with it, record other metrics, including  \textbf{Coverage} and \textbf{Surprisal}~\cite{vasilev2024replay}. All metrics are available in our publicly released dataset. 

\paragraph{Models:} We primarily focus on the SASRec model, as it is more commonly used in practice and frequently outperforms Bert4Rec~\cite{klenitskiy2023turning,mezentsev2024scalable}. However, to demonstrate the efficiency of \cce{} and \ccem{} for other architectures, we also present the results of applying these loss functions to the training of BERT4Rec in \S\ref{sec: efficiency_cce_ccem_for_Bert4Rec}

\paragraph{Baselines:} We analyze the efficiency of \cce{} and \ccem{} for 
SASRec by comparing metrics and computational resources against \ce{} and 
\cem{}, alternative 
sampling strategy such as SCE, and 
\bce{} . For time-memory comparisons, 
hyperparameters ($bs$, $sl$, $ns$) were selected to maximise NDCG@10 within 
a fixed GPU memory budget. We note that the best-performing configuration 
does not always correspond to full memory utilisation, in several cases, 
moderate hyperparameter values outperformed settings that pushed the memory 
budget to its limit, consistent with the saturation effects observed in 
\S\ref{sec:scaling}.

\subsection{Training Details}

For all datasets, we trained models on a single Nvidia A100 GPU. Prior to our experiments, we tuned the number of self-attention blocks, attention heads, and hidden size. As a result, the model configuration was set to two self-attention blocks with two attention heads and a hidden size $D$ of $256$. The models were trained using the Adam optimizer with a learning rate of $1e-3$. The training was performed with mixed-precision (float16) using the PyTorch Lightning trainer~\cite{micikevicius2017mixed}. 






\section{Results}
\label{sec:scaling}
In the following, we present the results of our computational experiments and analyze them to address the questions  \textbf{Q1} and \textbf{Q2} posed in \S\ref{sec:introduction} and evaluate the performance of \cce{} and \ccem{}. Specifically, (1) \S\ref{sec:scaling} contains the results of experiments carried out to study the effect of training hyperparameters on SASRec performance, where, among other insights, we demonstrate that sometimes increasing $ns$ does not lead to an improvement in the accuracy of the model; therefore, sometimes \cem{} should be preferred over \ce{}. (2) Further, we compare the training efficiency of the SASRec model using \cce{} and \ccem{} over \ce{}, \cem{} in  \S\ref{sec: comparative_analysis_ce_cem_cce_ccem}. The results of this section further highlight the effectiveness of negative sampling in \sr{}. (3) Since uniform negative sampling is not the only possible sampling strategy in \S\ref{sec: comparison_sampling_strategies}, we study other negative sampling strategies. (4) Finally, we extend our empirical analysis to the Bert4Rec model in \S\ref{sec: efficiency_cce_ccem_for_Bert4Rec}.

\subsection{Optimal Accuracy-Memory Trade-off}

\textbf{The first research question} is how to determine the optimal configuration of training hyperparameters given the memory constraints and how specific memory-related hyperparameters affect performance. We investigate how batch size (\textit{bs}), the number of negatives (\textit{ns}), and sequence length (\textit{sl}) affect the NDCG@10 metric of the SASRec model after training. We used a fixed grid of configurations, evaluating approximately \textbf{$170$} points per dataset, resulting in a total of \textbf{$997$} runs across all datasets. While the grid does not provide a fully uniform coverage of the hyperparameter space, the analysis can still be interpreted under the assumption of reasonably even exploration~\cite{scarlett2017lower,srinivas2010gaussian}.

To quantify the relative importance of the training hyperparameters, we fit an 
additive power law model independently for each dataset:
\[
NDCG_{\mathrm{norm}} = A + B \cdot sl^{SL} + C \cdot ns^{NS} + D \cdot bs^{BS}.
\]
Power law models of this form are commonly used to characterise how performance 
scales with training parameters~\cite{zhang2024scaling, kaplan2020scaling}. 
We adopt the additive formulation for its interpretability, each term isolates 
the marginal contribution of a single hyperparameter, and importance can be 
estimated directly from the observed range of each term.
Here, \(A\) is the baseline quality level for a dataset. The coefficients \(B\), \(C\), and \(D\) show how strongly \(sl\), \(ns\), and \(bs\) contribute to the predicted NDCG@10. Their sign shows whether the contribution is positive or negative, and their magnitude affects how large this contribution is.

The exponents \(SL\), \(NS\), and \(BS\) show how the effect changes when the 
corresponding hyperparameter grows. If an exponent is close to zero, the model 
is weakly sensitive to this hyperparameter; if its absolute value is larger, the 
effect changes more strongly. 

Results are presented in Table~\ref{tab:scaling_law_coefficients}. The 
per-dataset fits achieve $R^2 \geq 0.87$ on five out of six datasets, indicating 
a good fit of the additive power law model. The exception is MovieLens-20M 
($R^2 = 0.4149$), suggesting that the relationship between hyperparameters and 
quality is more complex or less monotone on this dataset, possibly due to its 
comparatively small catalog size. The last row reports a single 
model fitted jointly across all datasets, yielding $R^2 = 0.3459$, which 
confirms that hyperparameter importance is dataset-dependent and a single 
universal fit is insufficient.

\begin{table}[ht]
  \centering
  \caption{Scaling-law coefficients for dataset-specific fits of $NDCG = A + B \cdot sl^{SL} + C \cdot ns^{NS} + D \cdot bs^{BS}$. The last row reports the pooled fit over all datasets.}
  \label{tab:scaling_law_coefficients}
  \begin{tabular}{lrrrrrrrr}
    \toprule
    Dataset & $A$ & $B$ & $C$ & $D$ & $SL$ & $NS$ & $BS$ & $R^2 \uparrow$ \\
    \midrule
    Zvuk & 0.140091 & -0.521097 & 0.629307 & -1.883465 & -0.187984 & 0.087749 & -0.559963 & 0.9505 \\
    Megamarket & -0.101698 & -0.874397 & -1.340647 & 1.046564 & -0.292152 & -0.121650 & 0.102447 & 0.9663 \\
    30Music & 1.878231 & -1.975850 & -1.334797 & -1.855989 & -0.680028 & -0.248331 & -0.231238 & 0.8762 \\
    Beauty & -1.476963 & -0.084890 & -2.235147 & 1.490336 & -0.090727 & -0.444813 & 0.082969 & 0.6814 \\
    Gowalla & 1.082626 & -0.823191 & -0.814293 & -2.647282 & -0.482573 & -0.683141 & -0.541765 & 0.9249 \\
    Movielens-20m & 1.733849 & -0.977900 & -3.950294 & -2.782997 & -0.041238 & -1.000000 & -1.000000 & 0.4149 \\
    \midrule
    \textbf{For all datasets:} & 1.146648 & -1.159658 & -4.605289 & -1.463858 & -1.000000 & -0.840639 & -0.303288 & 0.3459 \\
    \bottomrule
  \end{tabular}
\end{table}

Since the final effect depends on both the coefficient and the power term, 
comparing \(B\), \(C\), and \(D\) directly would be misleading. To obtain a 
meaningful measure of each hyperparameter's contribution, we evaluate the range 
of each full term over the observed hyperparameter values within a dataset. 
Specifically, for $x \in \{sl, ns, bs\}$, let $term_{sl} = B \cdot sl^{SL}$, 
$term_{ns} = C \cdot ns^{NS}$, and $term_{bs} = D \cdot bs^{BS}$. The 
importance of hyperparameter $x$ is then defined as its term's range as a 
fraction of the total range across all three terms, when fitted parameters are fixed:
\[
imp_x =
\frac{\max(term_x)-\min(term_x)}
{\sum_{z \in \{sl,ns,bs\}}\left(\max(term_z)-\min(term_z)\right)}.
\]
This normalisation ensures that importance scores sum to one and are directly 
comparable across hyperparameters and datasets. 
\begin{table}[ht]
  \centering
  \caption{Scaling-law fit quality and hyperparameter importance. Importance is computed from the contribution range of each scaling-law term over the observed hyperparameter values for each dataset.}
  \label{tab:scaling_law_importance}
  \begin{tabular}{lrrrrl}
    \toprule
    Dataset & $R^2$ & $sl$ & $ns$ & $bs$ & Dominant \\
    \midrule
    Movielens-20m & 0.2462 & \textbf{37.6}\% & 33.5\% & 29.0\% & $sl$ \\
    Beauty & 0.6612 & 2.0\% & 35.8\% & \textbf{62.3}\% & $bs$ \\
    Gowalla & 0.8355 & 28.0\% & 12.3\% & \textbf{59.7}\% & $bs$ \\
    Zvuk & 0.9075 & 18.2\% & \textbf{49.2}\% & 32.5\% & $ns$ \\
    30Music & 0.8285 & 18.4\% & 34.6\% & \textbf{47.0}\% & $bs$ \\
    Megamarket & 0.9588 & 17.3\% & 25.9\% & \textbf{56.8}\% & $bs$ \\
    \bottomrule
  \end{tabular}
\end{table}
Table~\ref{tab:scaling_law_importance} shows that batch size is the dominant 
hyperparameter on four out of six datasets, followed by the number of negative 
samples, while sequence length is the least influential in most settings and can 
be reduced first when memory is constrained. No single hyperparameter 
universally dominates, however: on Zvuk, negative sampling is the strongest 
driver, and on MovieLens-20M all three contribute roughly equally. This 
suggests that under a fixed memory budget, the optimal strategy is not to 
maximise one hyperparameter at the expense of the others, but rather to balance 
all three.

To further guide practitioners, we derive the optimal $bs/ns$ allocation under 
a fixed joint memory budget $M = bs \cdot ns$ via a Lagrangian analysis of the 
fitted power law model (Appendix~\ref{app:scalinglaws}). Unlike the importance 
analysis, which quantifies the marginal value of each hyperparameter in 
isolation, the Lagrangian addresses how a given budget should be split between 
the two. The results confirm that the optimal ratio is dataset-dependent and 
should be computed from the fitted coefficients for each setting, The derived 
ratios are best treated as initial estimates rather than universal constants.

\begin{figure*}[ht]
  \centering
  \includegraphics[width=0.98\textwidth]{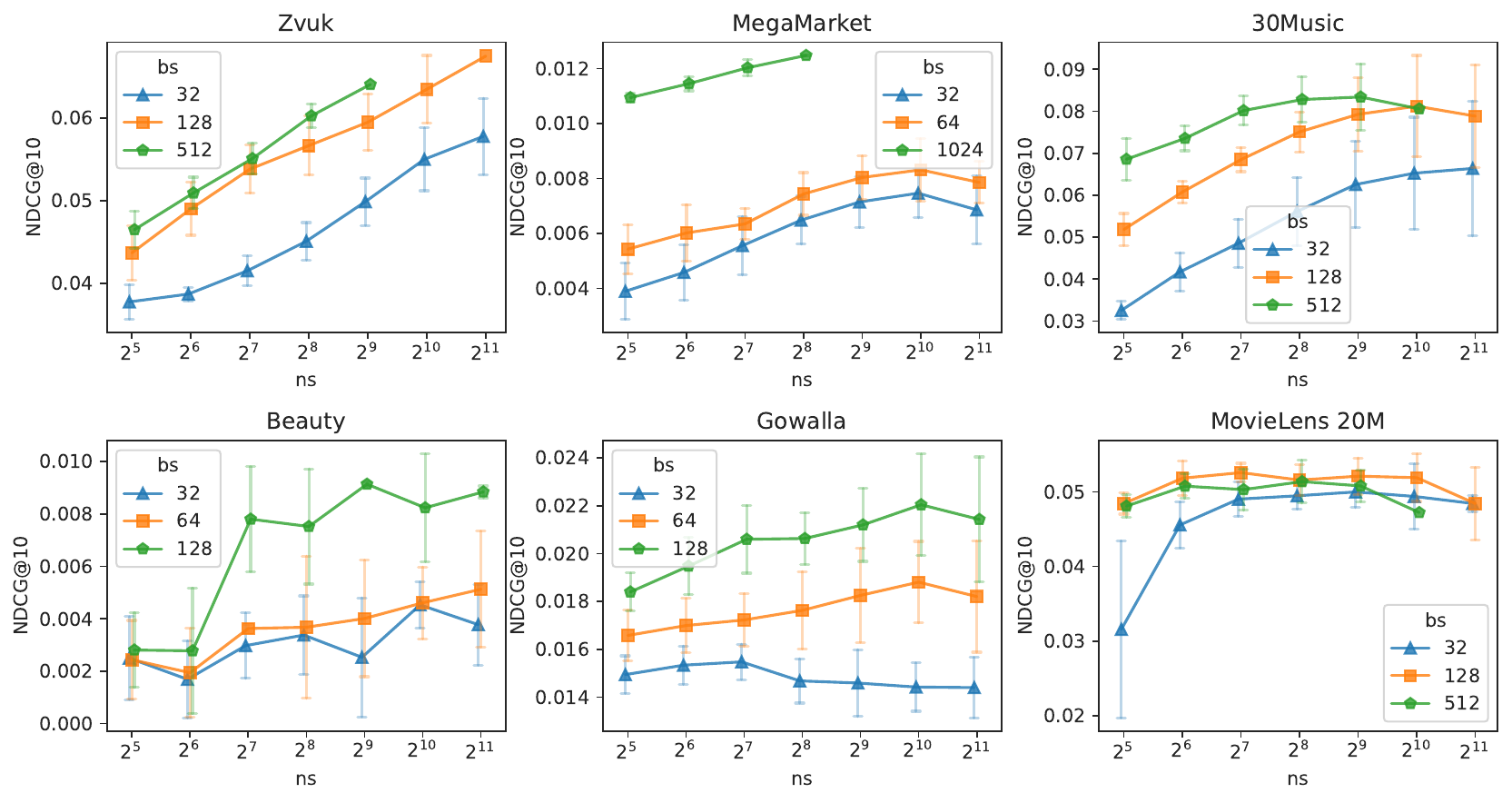}
\caption{This figure presents aggregated results from Figure~\ref{fig:negative_memory}, averaging across sequence lengths. Error bars indicate standard deviations. Colors represent batch sizes ($bs$). Performance increases with negative sample size ($ns$) for some datasets, while saturating for others (Movielens-20M, Gowalla). Batch size scaling shows no universal pattern: larger batches generally improve performance but saturate for 30Music, while smaller batches (128 vs 512) yield better results for Movielens-20M}
  \label{fig:negative}
  \Description{description}
\end{figure*}
\textbf{Effect of negative sampling on NDCG.} Figure~\ref{fig:negative} 
presents the relationship between $ns$ and NDCG@10 for the \cem{} loss. 
In most cases, the metric increases with $ns$ up to a certain point, after 
which it stabilizes or slightly decreases. The decrease is most prominent for 
Gowalla, likely due to overfitting, while for MegaMarket the influence of $ns$ 
is less pronounced throughout. Minor variations with $sl$ confirm the low 
sensitivity to sequence length observed in the power law analysis above.

\begin{figure}[!t]
  \centering
  \includegraphics[width=0.70\textwidth]{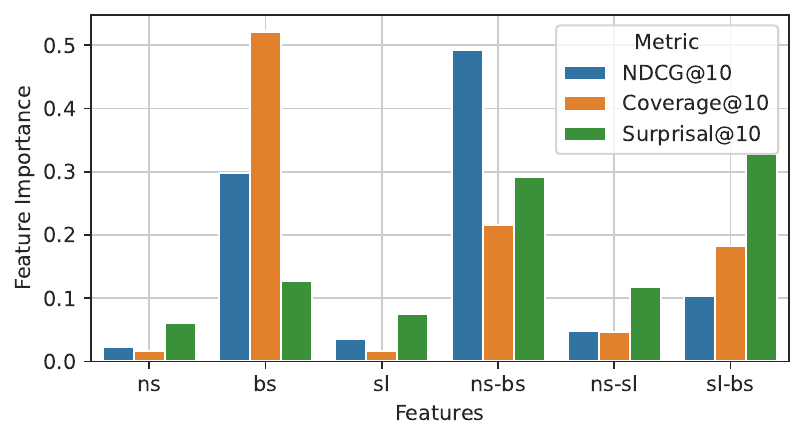}
  \caption{Feature importance (dataset average) from Random Forest Regression predicting NDCG@10, Coverage@10, and Surprisal@10. $ns-bs$, $ns-sl$, $sl-bs$ denote pairwise products as dependent variables.}
  \label{fig:importance}
  \Description{A bar plot showing Random Forest feature importance averaged across datasets for recommendation metrics. Batch size, number of negative samples, and their pairwise interactions have the largest predictive importance.}
\end{figure}

\textbf{Should we scale several dimensions simultaneously?} To complement the 
power law analysis, we train a Random Forest regressor using $bs$, $ns$, $sl$, 
and their pairwise products as features to predict NDCG@10, Coverage@10, and 
Surprisal@10. Models are constructed separately for each dataset and achieve 
$R^2 > 0.95$, confirming that the hyperparameters explain most of the variance 
in model quality. Figure~\ref{fig:importance} reports the average feature 
importance across datasets. $bs$ is the single most important feature for 
NDCG@10, consistent with Table~\ref{tab:scaling_law_importance}, while the 
high importance of the $ns$-$bs$ interaction confirms that the two should be 
scaled jointly rather than independently. $sl$ is individually weak but 
contributes through its interaction with $bs$.

Taken together, these results support the recommendation from the power law 
analysis: prioritise $bs$ and $ns$, scale them jointly, and reduce $sl$ first 
when memory is constrained. For large-catalog datasets such as Gowalla, 
30Music, Zvuk, and Megamarket, increasing all hyperparameters improves 
accuracy, but the associated memory costs make a memory-efficient training 
approach essential. Finally, we note that the comparatively low importance of 
$sl$ should be interpreted with care, as it may reflect properties of the 
specific datasets or model architecture considered here rather than a 
general conclusion.




\begin{table*}[th]
\centering
\small
\caption{Comparison of \ce{} and \cce{}. Hyperparameters such as $bs, sl$, and $ns$ were chosen based on our initial pool of scaling experiments for \ce{} and \cem{}, respectively. \cce{} and \ccem{} were trained using the same settings. We report NDCG@10, allocated memory (GB) and time per epoch (seconds). Bold values indicate best performance per dataset. \(\Delta\): Relative difference compared to \ce{} and \cem{}, green denotes \cce{} and \ccem{} superiority.}
\label{tab:ce_vs_cce}
\scriptsize
\resizebox{\textwidth}{!}{
\begin{tabular}{llccccccccc}
    \hline
    \textbf{Dataset} & \textbf{Parameters} 
    & \multicolumn{2}{c}{\textbf{NDCG@10}}
    & \multicolumn{2}{c}{\textbf{Mem, GB}}
    & \multicolumn{2}{c}{\textbf{Time, s}}
    & \textbf{$\Delta$ NDCG, \%}
    & \textbf{$\Delta$ Mem, \%}
    & \textbf{$\Delta$ Time, \%} \\
    \cmidrule(lr){3-4} \cmidrule(lr){5-6} \cmidrule(lr){7-8}
    & \textbf{(bs/sl)} & \ce{} & \cce{} & \ce{} & \cce{} & \ce{} & \cce{} & & & \\
    \hline
    Movielens-20m & 256/512
        & \textbf{0.0510} & 0.0505
        & 19.3 & \textbf{2.5}
        & 49.9 & \textbf{37.0}
        & 0.0\%
        & \textcolor{green!50!black}{↓ 87.0\%}
        & \textcolor{green!50!black}{↓ 25.9\%} \\
    Beauty & 1024/32
        & \textbf{0.0114} & 0.0104
        & 51.7 & \textbf{2.7}
        & 132.0 & \textbf{80.3}
        & \textcolor{red}{↓ 8.6\%}
        & \textbf{\textcolor{green!50!black}{↓ 97.2\%}}
        & \textcolor{green!50!black}{↓ 39.2\%} \\
    Gowalla & 256/64
        & 0.0225 & \textbf{0.0227}
        & 47.7 & \textbf{1.7}
        & 57.5 & \textbf{32.6}
        & 0.0\%
        & \textcolor{green!50!black}{↓ 96.4\%}
        & \textcolor{green!50!black}{↓ 43.3\%} \\
    Zvuk & 64/96
        & 0.0797 & \textbf{0.0803}
        & 65.2 & \textbf{4.4}
        & 1766.1 & \textbf{509.2}
        & 0.0\%
        & \textcolor{green!50!black}{↓ 93.3\%}
        & \textbf{\textcolor{green!50!black}{↓ 71.2\%}} \\
    30Music & 32/128
        & \textbf{0.0891} & 0.0841
        & 69.8 & \textbf{4.4}
        & 252.4 & \textbf{118.8}
        & \textcolor{red}{↓ 5.6\%}
        & \textcolor{green!50!black}{↓ 93.7\%}
        & \textcolor{green!50!black}{↓ 52.9\%} \\
    Megamarket & 16/128
        & \textbf{0.0052} & 0.0038
        & 64.7 & \textbf{15.7}
        & 4054.3 & \textbf{2936.8}
        & \textcolor{red}{↓ 26.6\%}
        & \textcolor{green!50!black}{↓ 75.7\%}
        & \textcolor{green!50!black}{↓ 27.6\%} \\
    \hline
    \textbf{Dataset} & \textbf{Parameters}
    & \multicolumn{2}{c}{\textbf{NDCG@10}}
    & \multicolumn{2}{c}{\textbf{Mem, GB}}
    & \multicolumn{2}{c}{\textbf{Time, s}}
    & \textbf{$\Delta$ NDCG, \%}
    & \textbf{$\Delta$ Mem, \%}
    & \textbf{$\Delta$ Time, \%} \\
    \cmidrule(lr){3-4} \cmidrule(lr){5-6} \cmidrule(lr){7-8}
    & \textbf{(bs/sl/ns)} & \cem{} & \ccem{} & \cem{} & \ccem{} & \cem{} & \ccem{} & & & \\
    \hline
    Movielens-20m & 64/512/1511
        & \textbf{0.0513} & 0.0488
        & 39.0 & \textbf{1.6}
        & 233.0 & \textbf{87.0}
        & \textcolor{red}{↓ 4.9\%}
        & \textbf{\textcolor{green!50!black}{↓ 95.9\%}}
        & \textcolor{green!50!black}{↓ 62.7\%} \\
    Beauty & 1024/32/511
        & \textbf{0.0123} & 0.0117
        & 11.7 & \textbf{2.7}
        & 36.8 & \textbf{23.0}
        & \textcolor{red}{↓ 5.1\%}
        & \textcolor{green!50!black}{↓ 87.2\%}
        & \textcolor{green!50!black}{↓ 37.5\%} \\
    Gowalla & 512/256/511
        & \textbf{0.0277} & 0.0276
        & 42.7 & \textbf{5.3}
        & 36.7 & \textbf{36.0}
        & 0.0\%
        & \textcolor{green!50!black}{↓ 87.6\%}
        & \textcolor{green!50!black}{↓ 23.4\%} \\
    Zvuk & 1024/32/255
        & \textbf{0.0593} & 0.0566
        & 56.0 & \textbf{6.9}
        & 58.0 & \textbf{35.0}
        & \textcolor{red}{↓ 4.5\%}
        & \textcolor{green!50!black}{↓ 87.7\%}
        & \textbf{\textcolor{green!50!black}{↓ 71.7\%}} \\
    30Music & 256/128/511
        & 0.1107 & \textbf{0.1132}
        & 47.6 & \textbf{3.6}
        & 49.6 & \textbf{18.6}
        & \textbf{\textcolor{green!50!black}{↑ 2.2\%}}
        & \textcolor{green!50!black}{↓ 92.4\%}
        & \textcolor{green!50!black}{↓ 62.5\%} \\
    Megamarket & 1024/32/255
        & \textbf{0.0117} & \textbf{0.0117}
        & 71.0 & \textbf{8.9}
        & 60.0 & \textbf{30.0}
        & 0.0\%
        & \textcolor{green!50!black}{↓ 87.5\%}
        & \textcolor{green!50!black}{↓ 50\%} \\
    \hline
\end{tabular}
}
\end{table*}
\begin{table*}[th]
\centering
\small
\caption{Comparison of NDCG@10, memory consumption, and epoch time for \cce{} and \ccem{} on the SASRec model using identical training hyperparameters, except \(ns\) for \ccem{}. Bold values indicate the best performance per dataset. \(\Delta\): Relative difference compared to \cce{}, green denotes \ccem{} superiority.}
\label{tab:cce_acc_mem}
\renewcommand{\arraystretch}{1.2}
\resizebox{\textwidth}{!}{
\begin{tabular}{llccccccccc}
\hline
\textbf{Dataset} & \textbf{Parameters}
& \multicolumn{3}{c}{\textbf{NDCG@10}}
& \multicolumn{3}{c}{\textbf{Mem, GB}}
& \multicolumn{3}{c}{\textbf{Time, s}} \\
\cmidrule(lr){3-5} \cmidrule(lr){6-8} \cmidrule(lr){9-11}
& \textbf{(bs/sl/ns)}
& \cce{} & \ccem{} & \textbf{$\Delta$, \%}
& \cce{} & \ccem{} & \textbf{$\Delta$, \%}
& \cce{} & \ccem{} & \textbf{$\Delta$, \%} \\
\hline
Movielens-20m & 256/512/1511
    & 0.0511 & \textbf{0.0523} & \textcolor{green!50!black}{↑ 2.3\%}
    & \textbf{3.4} & 3.5 & \textcolor{red}{↑ 2.9\%}
    & \textbf{43.0} & 52.0 & \textcolor{red}{↑ 20.9\%} \\

Beauty & 1024/32/511
    & 0.0104 & \textbf{0.0116} & \textcolor{green!50!black}{↑ 11.5\%}
    & 2.7 & 2.7 & \textcolor{red}{↑ 7.1\%}
    & 80.3 & \textbf{23.0} & \textcolor{green!50!black}{↓ 71.4\%} \\

Gowalla & 1024/320/511
    & 0.0241 & \textbf{0.0279} & \textcolor{green!50!black}{↑ 15.8\%}
    & 11.8 & \textbf{11.1} & \textcolor{red}{↑ 2.5\%}
    & 132.0 & \textbf{21.0} & \textcolor{green!50!black}{↓ 84.1\%} \\

Zvuk & 1024/128/2047
    & \textbf{0.0940} & 0.0852 & \textcolor{red}{↓ 10.6\%}
    & \textbf{10.6} & 32.3 & \textcolor{red}{↑ 47.4\%}
    & 550.0 & \textbf{308.0} & \textcolor{green!50!black}{↓ 44.0\%} \\

30Music & 256/720/4095
    & 0.1393 & \textbf{0.1436} & \textcolor{green!50!black}{↑ 3.1\%}
    & \textbf{9.4} & 58.4 & \textcolor{red}{↑ 261.7\%}
    & 389.0 & \textbf{277.0} & \textcolor{green!50!black}{↓ 28.8\%} \\

Megamarket & 5120/320/1023
    & \textbf{0.0181} & 0.0168 & \textcolor{red}{↓ 7.2\%}
    & \textbf{55.0} & 78.0 & \textcolor{red}{↑ 100.0\%}
    & 2248.0 & \textbf{262.0} & \textcolor{green!50!black}{↓ 88.3\%} \\
\hline
\end{tabular}
}
\end{table*}

\begin{table*}[th]
\centering
\small
\caption{Comparison of Coverage@10 and Surprisal@10 for \cce{} and \ccem{} on the SASRec model using identical training hyperparameters, except \(ns\) for \ccem{}. Bold values indicate the best performance per dataset. \(\Delta\): Relative difference compared to \cce{}, green denotes \ccem{} superiority.}
\label{tab:cce_coverage_surprisal}
\renewcommand{\arraystretch}{1.2}
\resizebox{0.75\textwidth}{!}{
\begin{tabular}{llcccccc}
\hline
\textbf{Dataset} & \textbf{Parameters}
& \multicolumn{3}{c}{\textbf{Coverage@10}}
& \multicolumn{3}{c}{\textbf{Surprisal@10}} \\
\cmidrule(lr){3-5} \cmidrule(lr){6-8}
& \textbf{(bs/sl/ns)}
& \cce{} & \ccem{} & \textbf{$\Delta$, \%}
& \cce{} & \ccem{} & \textbf{$\Delta$, \%} \\
\hline
Movielens-20m & 256/512/1511
    & 0.2102 & \textbf{0.2429} & \textcolor{green!50!black}{↑ 15.6\%}
    & 0.0793 & \textbf{0.0801} & \textbf{\textcolor{green!50!black}{↑ 1.0\%}} \\

Beauty & 1024/32/511
    & 0.1155 & \textbf{0.1813} & \textbf{\textcolor{green!50!black}{↑ 57.0\%}}
    & \textbf{0.5172} & 0.5069 & \textcolor{red}{↓ 2.0\%} \\

Gowalla & 1024/320/511
    & 0.2021 & \textbf{0.2770} & \textcolor{green!50!black}{↑ 37.1\%}
    & \textbf{0.6447} & 0.6276 & \textcolor{red}{↓ 2.7\%} \\

Zvuk & 1024/128/2047
    & 0.1401 & \textbf{0.1490} & \textcolor{green!50!black}{↑ 6.3\%}
    & \textbf{0.3547} & 0.3486 & \textcolor{red}{↓ 1.7\%} \\

30Music & 256/720/4095
    & 0.1203 & \textbf{0.1241} & \textcolor{green!50!black}{↑ 3.2\%}
    & 0.5783 & \textbf{0.5834} & 0\% \\

Megamarket & 5120/320/1023
    & \textbf{0.1705} & 0.1158 & \textcolor{red}{↓ 32.1\%}
    & \textbf{0.5466} & 0.4941 & \textcolor{red}{↓ 9.6\%} \\
\hline
\end{tabular}
}
\end{table*}

\subsection{\texorpdfstring{Comparative Analysis: \ce{}/\cem{} vs. \cce{}/\ccem{}}{Comparative Analysis: CE/CE- vs. CCE/CCE-}}

\label{sec: comparative_analysis_ce_cem_cce_ccem}
In this subsection, we analyze the computational efficiency during SASRec training with  \ce{}/\cce{}, \cem{}/\ccem{}, and \cce{}/\ccem{}. Also, we compare the model's performance metrics.

Although \ce{}/\cce{} and \cem{}/\ccem{} are mathematically equivalent, their implementation code for GPU computation differs. This leads to variations in memory footprint, computational time, and numerical deviations, motivating a detailed comparison.


In Table~\ref{tab:ce_vs_cce} and Figure~\ref{fig:best_metrics}, we evaluate the \textbf{memory footprint} and \textbf{training speed} (time per epoch) of SASRec under different loss configurations. The hyperparameters \(bs\), \(sl\), and \(ns\) were optimized for \ce{}/\cem{} and \cce{}/\ccem{} based on the scaling experiments in \S\ref{sec:scaling}, ensuring a fair comparison of their computational requirements.  

\textbf{Memory consumption significantly contracts across all datasets for the \cce{}}, indicating that the memory footprint is mainly due to the materialization of logits for computing the \ce{} loss. 
The time per epoch also decreases by a minimum of 25.9\%. This suggests that the fusion of the log-sum-exp operation with matrix multiplication and gradient filtering implemented in the \cce{} kernels enhances the efficiency of computing the \ce{} loss.

\textbf{Significant reduction of computational resources occurs for \ccem{} compared to \cem{}. }
The memory footprint decreases by no less than 87\%  for all datasets. The time per epoch also reduces by more than 23\%, achieving 73\% for \textit{Zvuk}.

\textbf{Deviation in metrics:} Although resource savings are evident, NDCG@10 performance slightly decreases on some datasets when training with \cce{} or \ccem{} compared to \ce{} and \cem{}. The only exception is the \textit{30Music} dataset, where \ccem{} shows an improvement over \cem{}.

Since the loss function formulation in \cce{} and \ccem{} is simply a logarithmic decomposition of the $\mathrm{LogSoftmax}$ function used in the PyTorch implementation of \ce{} and \cem{}, the observed metrics discrepancies may arise from minor numerical inaccuracies during intermediate computations in Triton kernels or from loss scaling in mixed-precision training. However, this issue requires further investigation, which will be addressed in future work.



\begin{table*}[th]
\centering
\small
\caption{Results for \ccem{} with popularity sampling on the SasRec model using identical hyperparameters as for global uniform sampling}
\label{tab:cce_popularity}
\renewcommand{\arraystretch}{1.2}
\resizebox{0.85\textwidth}{!}{
\begin{tabular}{llccccc}
  \hline
    \textbf{Dataset} & \textbf{Parameters (bs/sl/ns)} & NDCG@10 & Coverage@10 & Surprisal@10 & Mem (GB) & Time (s) \\
 \hline
    Movielens-20m & 256 / 512 / 1511 & 0.0235 & 0.2221 & 0.1337 & 6.2 & 111.0 \\
    Beauty & 1024 / 32 / 511 & 0.0066 & 0.5181 & 0.7098 & 5.7 & 82.9 \\
    Gowalla & 1024 / 320 / 511 & 0.0200 & 0.4540 & 0.7521 & 67.9 & 86.9 \\
   \hline
\end{tabular}
}
\end{table*}

\begin{table*}[h]
\centering
\small
\caption{Results for the comparison of \ccem{} loss with global uniform and popularity sampling for the SASRec model. \(\Delta\): Relative difference compared to popularity sampling and global uniform sampling, green denotes global uniform sampling superiority.}
\label{tab:cce_comparison_global_popularity}
\renewcommand{\arraystretch}{1.2}
\resizebox{0.85\textwidth}{!}{
\begin{tabular}{lcccccc}
\hline
Dataset & \textbf{(bs/sl)} &  \textbf{$\Delta$ NDCG, \%} & \textbf{$\Delta$ Coverage, \%} & \textbf{$\Delta$ Surprisal, \%} & \textbf{$\Delta$ Mem, \%} & \textbf{$\Delta$ Time, \%} \\
\hline
Movielens-20m   & 256  / 512 &  \textcolor{green!50!black}{↑ 55.1\%}  & \textcolor{green!50!black}{↑ 8.6\%} & \textcolor{red}{↓ 66.9\%} & \textcolor{green!50!black}{↓ 77.1\%} & \textcolor{green!50!black}{↓ 113.5\%} \\
Beauty          & 1024 / 32  &  \textcolor{green!50!black}{↑ 43.1\%} & \textcolor{red}{↓ 185.8\%} & \textcolor{red}{↓ 40.0\%} & \textcolor{green!50!black}{↓ 111.1\%} & \textcolor{green!50!black}{↓ 260.4\%} \\
Gowalla         & 1024 / 320 &  \textcolor{green!50!black}{↑ 28.3\%} & \textcolor{red}{↓ 63.9\%} & \textcolor{red}{↓ 19.8\%} & \textcolor{green!50!black}{↓ 511.7\%} & \textcolor{green!50!black}{↓ 313.8\%}  \\ 

\hline
\end{tabular}
}
\end{table*}

\subsection{\texorpdfstring{Reallocating 
memory savings to increase performance with \cce{} and \ccem{}}{Reallocating memory savings to increase performance with CCE and CCE-}}

The results discussed in the previous section indicate that \cce{} and \ccem{} significantly reduce the memory consumption required for training SASRec. This reduction provides an opportunity to enhance model performance by increasing batch size ($bs$), sequence length ($sl$), and number of samples ($ns$) in the case of \ccem{}, implementing insights from the \S\ref{sec:scaling}. 
In these experiments, we set identical training hyperparameters for both \cce{} and \ccem{}. Tables \ref{tab:cce_acc_mem}, \ref{tab:cce_coverage_surprisal} present the values of the NDSG@10, Coverage@10, and Surprisal@10 metrics for all datasets, along with the allocated memory and time per epoch required for training.

\textbf{Improvement in model performance across all datasets is observed} when comparing results in Tables~\ref{tab:ce_vs_cce}, ~\ref{tab:cce_acc_mem}, \ref{tab:cce_coverage_surprisal} and in Figure~\ref{fig:best_metrics}, except for smaller \textit{Beauty}, where adjustments to the training parameters do not yield an increase in metrics. 
The most significant improvements are observed in \textit{Megamarket}, which shows a 54.7\% increase, followed by \textit{30Music} with a 27\% increase, and \textit{Zvuk} with a 17.1\% increase. These datasets are characterized by extensive catalogs, exceeding 800K items. 
Conversely, for datasets \textit{Beauty}, \textit{Gowalla}, \textit{Movielens-20m}, which have item catalogs not exceeding 310K items, applying \cce{} and \ccem{} for model training does not lead to considerable improvement in their performance. 
Thus, using the \cce{} and \ccem{} for training models with large final layers improves performance, while for smaller catalogs main benefit is training acceleration.

\textbf{Negative sampling during training with the \ccem{}  accelerates training compared to \cce{} across datasets} except \textit{Movielens-20m} (ML20M), which includes only 15K items. For example, using 1023 negative samples for \textit{Megamarket} results in a time per epoch reduction of 88.3\% compared to \cce{}, while using 4095 negative samples for 30Music leads to only a 28.8\% acceleration. Additionally, restricting the item catalog during scoring may degrade model performance, particularly for datasets with large item catalogs like \textit{Megamarket} and \textit{Zvuk}. In contrast, for datasets with fewer items, such as \textit{30Music}, \textit{Gowalla}, and \textit{Beauty}, training with \ccem{} can yield improvements in model metrics compared to \cce{}.

\textbf{What is better \ccem{} over \cce{}?} We show that \ccem{} is faster and outperforms \cce{} on four out of six datasets due to negative sampling, however, it requires more memory to store a matrix of negative sample indices compared to \cce{}. 

\subsection{Gradient Filtering with \cce{}}
Here we attempt to answer \textbf{Q2} posed in \S\ref{sec:introduction}.
To quantify the impact of pruning near-zero gradients, we conducted experiments using \ccem{} across six benchmark datasets. The training hyperparameters were kept consistent with those used in the \cce{} and \ccem{} comparison experiments. To analyze gradient sparsity, we performed a series of experiments where the threshold for zeroing out gradients was incrementally increased in each run.

\begin{figure}[ht]
  \centering
  \includegraphics[width=0.95\textwidth]{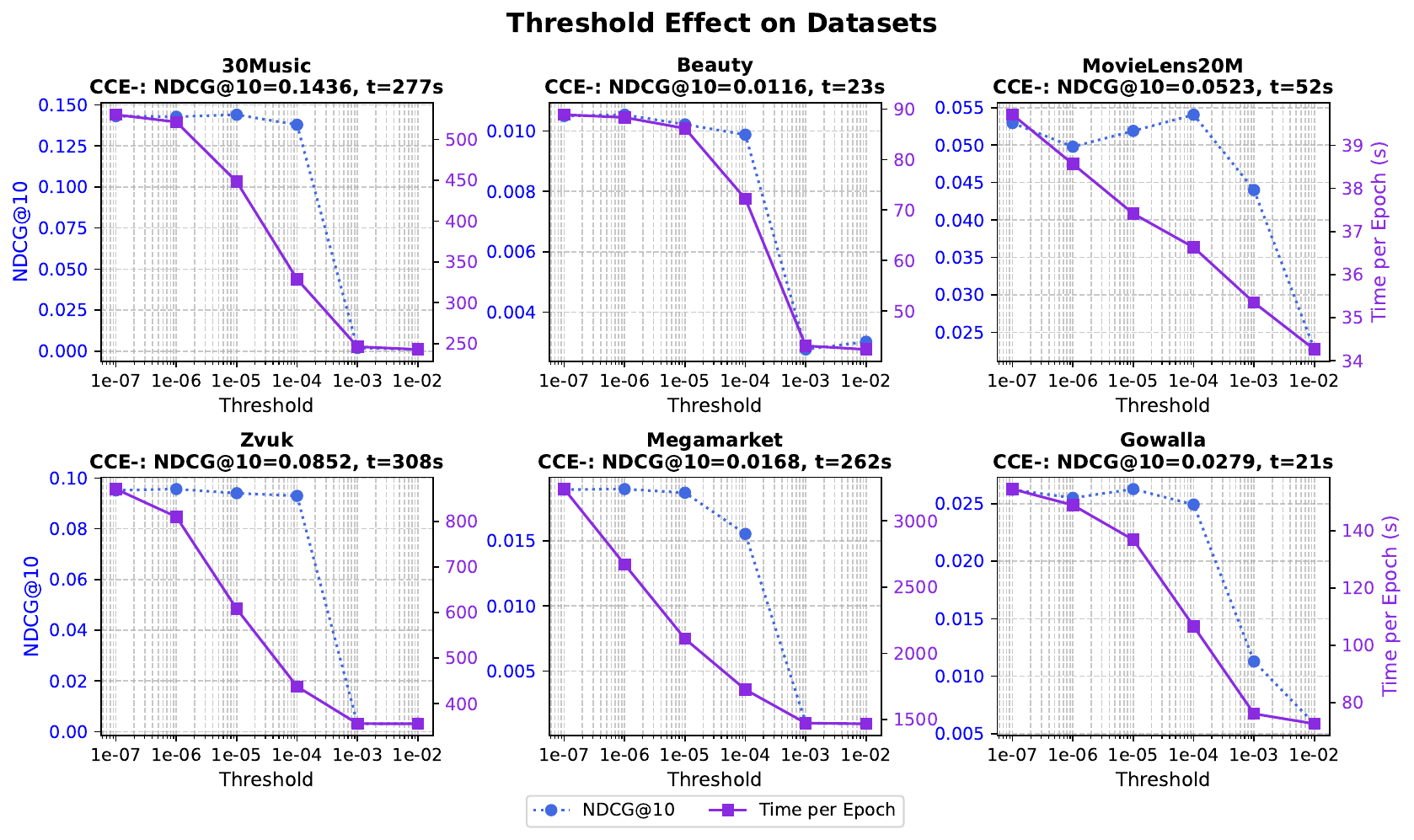}
\caption{Effect of the gradient-filtering threshold \texttt{filter\_eps} on SASRec trained with CCE. In each subplot, the curves show NDCG@10 (left axis) and training time per epoch (right axis) for different threshold values, while the subplot title reports the corresponding CCE$-$ result on the same dataset.}
\label{fig:sparse_grads_cce}
\Description{description}
\end{figure}

Figure~\ref{fig:sparse_grads_cce} shows training time per epoch and NDCG@10 metric of the model. We can observe two trends:  (1) \textbf{Performance stability}: Model accuracy degrades only when the threshold exceeds \(1 \times 10^{-4} - 1 \times 10^{-3}\).  (2) \textbf{Efficiency gains}: 
The training time per epoch consistently enhances as computations of near-zero gradients are omitted.

These results suggest that gradients close to zero contribute minimally to model updates during later training stages, enabling safe pruning for computational savings. Moreover, for \textit{Movielens-20M} gradient pruning induces regularization that leads to improved NDCG@10 metrics when the pruning threshold increases from $1 \times 10^{-6} - 1$ to $1 \times 10^{-4} - 1$. This indicates that for this dataset, there are many noisy gradients that complicate training. 

Since \cce{} with gradient pruning enables accelerating the training process, we compare the time per epoch and the accuracy of models trained with \cce{} and \ccem{}. The training hyperparameters matched those specified in Table \ref{tab:cce_acc_mem}. 

Based on the results shown in Figure\ref{fig:sparse_grads_cce}, it can be concluded that \ccem{} achieves NDCG@10 scores comparable to those of \cce{}. At that training with \ccem{} requires the least time per epoch on five of the six datasets, with the exception of \textit{Movielens-20m}. For the \textit{Beauty}, \textit{Megamarket}, and \textit{Gowalla} datasets, \ccem{} achieves over 3x acceleration over \cce{} despite of using gradient filtering at a threshold of $1\times10^{-4}$. Under these conditions, the NDCG@10 score for \ccem{} is higher across all three datasets, with the smallest observed improvement, an 8\% increase, occurring in the Megamarket dataset. 

For the \textit{Movielens-20m} dataset, training with \cce{} is approximately 35\% faster per epoch than \ccem{}, even without gradient filtering. This dataset features the smallest item catalog among the considered datasets, comprising approximately 15K items \ref{tab:datasets}.
At that, \ccem{} requires $1511$ negative samples to reach an NDCG@10 score comparable to that of \ccem{}. The large number of negative samples relative to the catalog size causes significant overhead when computing the forward and backward passes in \ccem{} for the classification layer.

\subsubsection{Comparison Sampling Strategies}
\label{sec: comparison_sampling_strategies}
One of the main features of training with the \ce{} loss using negative sampling is the ability to employ diverse strategies for selecting negative samples. In this work, we primarily focus on the global uniform strategy, where negative samples are drawn uniformly across the entire dataset. 

As an alternative, we also consider popularity-based negative sampling, where negative samples are selected based on item popularity. In this case, the probability of selection increases with an item’s popularity, causing popular items to be chosen as negative samples more frequently.

Tables \ref{tab:cce_popularity}, \ref{tab:cce_comparison_global_popularity}  present data to compare metrics and computational resource usage for training with \ccem{} using global uniform and popularity-based sampling strategies. In terms of $\mathrm{NDSG@10}$ metric, global uniform sampling consistently outperforms the popularity-based approach across all considered datasets. However, popularity-based sampling shows significant gains in diversity-related metrics for the \textit{Gowalla} and \textit{Beauty} datasets. Compared to popularity-based sampling, using global uniform sampling leads to a reduction in $\mathrm{Coverage@10}$ by 185.8\% and 63.9 \%, and a reduction in $\mathrm{Surprisal@10}$
by 40\% and 19.8\%. These results align with observations that popularity-weighted sampling enhances exposure to long-tail items, improving recommendation diversity and novelty \cite{wang2024popularitylearning}.

The popularity sampling strategy demands higher memory consumption during training compared to the global uniform sampling strategy. This increased requirement is due to the implementation characteristics of \textbf{torch.multinomial}, which necessitates the materialization of a tensor representing the unnormalized probabilities for sampling negative items in popularity-based scenarios. Consequently, only datasets with relatively small item catalogs can be processed under this strategy.

Another approach to negative sampling is the Scalable Cross-Entropy (\sce{}) method ~\cite{mezentsev2024scalable}. This method approximates the \ce{} loss by employing a sampling procedure that computes logits only for elements with the largest absolute values, rather than for all items in the catalog. As a result, memory usage for the logit tensor is significantly reduced, and the loss computation is accelerated. 

Table \ref{tab:sasrec_sce} presents the results of the SASRec model trained with the \sce{} loss. Tables \ref{tab:sasrec_sce_ccem_comparison}, \ref{tab:sasrec_sce_cce_comparison} list the relative differences calculated for the metrics achieved by \cce{} and \ccem{}.

When comparing \sce{} with \ccem{}, \ccem{} achieves considerably higher $\mathrm{NDCG@10}$ scores on four out of six datasets, with the exception of \textit{Zvuk} and \textit{Megamarket}. However, \sce{} provides better $\mathrm{Suprisal@10}$ metrics for all datasets. 

Compared to \ccem{}, \sce{} requires more memory on the \textit{Megamarket} dataset. Therefore, training with SCE was conducted using a batch size of 2048.  In this case, \sce{} achieves $\mathrm{NDSG@10}$ score comparable with \ccem{}, which is applied for batch size 5120. 

For \cce{} $\mathrm{NDSG@10}$ values are higher than those of SCE across all datasets. On the \textit{Megamarket} \cce{} with batch size 5120 also outperforms SCE on  11.6 \% in terms of $\mathrm{NDSG@10}$.  Moreover, \cce{} requires significantly less memory during training, reducing memory usage by approximately 80\% on both the \textit{30Music} and \textit{Beauty} datasets.

The time per epoch for \sce{} is considerably lower than for \cce{} and \ccem{} on datasets with large item catalogs, such as \textit{Zvuk}, \textbf{30Music}, and \textit{Megamarket}. Therefore, this sampling approach can offer improved computational efficiency when training RecSys models with a large number of logits. Since \sce{} incorporates the \ce{} loss function, it could be further optimized in the future using our Triton kernels, which compute the CE loss with negative sampling by using only the indices of positive and negative samples. This eliminates the need to materialize logits, thereby reducing memory consumption during training.

\begin{table*}[th]
\centering
\small
\caption{Results for the \textbf{SCE} loss with the SASRec model}
\label{tab:sasrec_sce}
\renewcommand{\arraystretch}{1.2}
\resizebox{0.85\textwidth}{!}{
\begin{tabular}{lcccccc}
\hline
\textbf{Dataset} & \textbf{Parameters} & \multicolumn{5}{c}{\textbf{SCE}} \\
\cmidrule(lr){3-7}
& \textbf{(bs/sl)} & \textbf{NDCG@10} & \textbf{Coverage@10} & \textbf{Surprisal@10} & \textbf{Mem, GB} & \textbf{Time, s} \\
\hline
Movielens-20m  & 64   / 512 & 0.0477 & 0.1262 & 0.0670 & 3.6   & 34.2  \\
Movielens-20m  & 256  / 512 & 0.0396 & 0.1812 & 0.0856 & 5.4   & 69.5  \\
Beauty         & 1024 / 32  & 0.0086 & 0.2290 & 0.5746 & 4.9   & 167.8 \\
Gowalla        & 512  / 256 & 0.0205 & 0.2805 & 0.6676 & 7.9   & 14.8  \\
Gowalla        & 1024 / 320 & 0.0222 & 0.3463 & 0.6887 & 15.9  & 23.6  \\
Zvuk           & 1024 / 32  & 0.0781 & 0.0804 & 0.3725 & 10.1  & 34.0  \\
Zvuk           & 1024 / 128 & 0.0879 & 0.1109 & 0.3711 & 13.5  & 68.4  \\
30Music        & 256  / 128 & 0.1011 & 0.0912 & 0.5682 & 8.1   & 13.7  \\
30Music        & 256  / 720 & 0.1218 & 0.1065 & 0.5812 & 16.7  & 45.5  \\
Megamarket     & 1024 / 32  & 0.0124 & 0.0706 & 0.5278 & 14.5  & 36.4  \\
Megamarket     & 2048 / 320 & 0.0160 & 0.0988 & 0.5358 & 57.0  & 144.2 \\
\hline
\end{tabular}
}
\end{table*}

\begin{table*}[h]
\centering
\small
\caption{Results for the comparison of \textbf{SCE} loss with \ccem{} for the SASRec model. \(\Delta\): Relative difference compared to \textbf{SCE} and \ccem{}, green denotes \ccem{} superiority.}
\label{tab:sasrec_sce_ccem_comparison}
\renewcommand{\arraystretch}{1.2}
\resizebox{0.85\textwidth}{!}{
\begin{tabular}{lcccccc}
\hline
Dataset & \textbf{(bs/sl)} &  \textbf{$\Delta$ NDCG, \%} & \textbf{$\Delta$ Coverage, \%} & \textbf{$\Delta$ Surprisal, \%} & \textbf{$\Delta$ Mem, \%} & \textbf{$\Delta$ Time, \%} \\
\hline
Movielens-20m   & 256  / 512 &  \textcolor{green!50!black}{↑ 24.3\%}  & \textcolor{green!50!black}{↑ 25.9\%} & \textcolor{red}{↓ 6.9\%} & \textcolor{green!50!black}{↓ 54.3\%} & \textcolor{green!50!black}{↓ 33.7\%} \\
Beauty          & 1024 / 32  &  \textcolor{green!50!black}{↑ 25.9\%} & \textcolor{red}{↓ 26.3\%} & \textcolor{red}{↓ 13.4\%} & \textcolor{green!50!black}{↓ 81.4\%} & \textcolor{green!50!black}{↓ 629.5\%} \\
Gowalla         & 1024 / 320 &  \textcolor{green!50!black}{↑ 20.4\%} & \textcolor{red}{↓ 25.1\%} & \textcolor{red}{↓ 9.7\%} & \textcolor{green!50!black}{↓ 43.2\%} & \textcolor{green!50!black}{↓ 12.4\%}  \\
Zvuk            & 1024 / 128 &  \textcolor{red}{↓ 3.2\%} & \textcolor{green!50!black}{↑ 25.6\%} & \textcolor{red}{↓ 6.5\%} & \textcolor{red}{↑ 58.2\%} & \textcolor{red}{↑ 77.8\%} \\ 
30Music         & 256  / 128 &   \textcolor{green!50!black}{↑ 15.2\%}  & \textcolor{green!50!black}{↑ 14.2\%} & 0.0\% & \textcolor{red}{↑ 71.4\%} & \textcolor{red}{↑ 83.6\%} \\
Megamarket      & 2048 (5120\ccem{}) / 320 &  \textcolor{green!50!black}{↑ 4.8\%} &  \textcolor{green!50!black}{↑ 14.7\%}  &  \textcolor{red}{↓ 8.4\%} &  \textcolor{red}{↑ 26.9\%}  &  \textcolor{red}{↑ 45.0\%} \\

\hline
\end{tabular}
}
\end{table*}

\begin{table*}[h]
\centering
\small
\caption{Results for the comparison of \textbf{SCE} loss with \cce{} for the SASRec model. \(\Delta\): Relative difference compared to \textbf{SCE} and \ccem{}, green denotes \cce{} superiority.}
\label{tab:sasrec_sce_cce_comparison}
\renewcommand{\arraystretch}{1.2}
\resizebox{0.85\textwidth}{!}{
\begin{tabular}{lcccccc}
\hline
Dataset & \textbf{(bs/sl)} &  \textbf{$\Delta$ NDCG, \%} & \textbf{$\Delta$ Coverage, \%} & \textbf{$\Delta$ Surprisal, \%} & \textbf{$\Delta$ Mem, \%} & \textbf{$\Delta$ Time, \%} \\
\hline
Movielens-20m   & 256  / 512 &  \textcolor{green!50!black}{↑ 22.5\%}  & \textcolor{green!50!black}{↑ 13.8\%} & \textcolor{red}{↓ 7.9\%} & \textcolor{green!50!black}{↓ 58.8\%} & \textcolor{green!50!black}{↓ 61.6\%} \\
Beauty          & 1024 / 32  &  \textcolor{green!50!black}{↑ 17.3\%} & \textcolor{red}{↓ 98.3\%} & \textcolor{red}{↓ 11.1\%} & \textcolor{green!50!black}{↓ 81.5\%} & \textcolor{green!50!black}{↓ 109.0\%} \\
Gowalla         & 1024 / 320 &  \textcolor{green!50!black}{↑ 7.9\%} & \textcolor{red}{↓ 71.4\%} & \textcolor{red}{↓ 6.8\%} & \textcolor{green!50!black}{↓ 34.7\%} & \textcolor{red}{↑ 82.1\%}  \\
Zvuk            & 1024 / 128 &  \textcolor{green!50!black}{↑ 6.5\%} & \textcolor{green!50!black}{↑ 20.8\%} & \textcolor{red}{↓ 4.6\%} & \textcolor{green!50!black}{↓ 27.4\%} & \textcolor{red}{↑ 87.6\%} \\ 
30Music         & 256  / 720 &   \textcolor{green!50!black}{↑ 12.6\%}  & \textcolor{green!50!black}{↑ 11.5\%} &  \textcolor{red}{↓ 0.5\%} & \textcolor{green!50!black}{↓ 77.7\%} & \textcolor{red}{↑ 88.3\%} \\
Megamarket      & 2048 (5120\ccem{}) / 320 &  \textcolor{green!50!black}{↑ 11.6\%} &  \textcolor{green!50!black}{↑ 42.1\%}  &  \textcolor{green!50!black}{↑ 2.0\%} &  \textcolor{green!50!black}{↓ 3.6\%}  &  \textcolor{red}{↑ 93.6\%} \\

\hline
\end{tabular}
}
\end{table*}

\subsection{\texorpdfstring{Comparison of \cce{} and \ccem{} with \bce{}}{Comparison of CCE and CCE- with BCE}}
\label{sec: comparison_cce_ccem_with_bce_sasrec}

As outlined in the original paper \cite{kang2018self}, the SASRec model can be trained using the Binary Cross-Entropy (\bce{}) loss. To evaluate the effectiveness of our proposed methods, we conducted training SASRec with the \bce{} loss.

Table~\ref{tab:sasrec_bce} summarizes the performance of SASRec trained with the \bce{} loss. A comparison with models trained using the \ccem{} and \cce{} losses is presented in Tables~\ref{tab:sasrec_bce_ccem} and~\ref{tab:sasrec_bce_cce}. Under identical settings for batch size and sequence length, SASRec with the \bce{} loss achieves significantly lower NDCG@10 and Coverage@10 scores compared to both \ccem{} and \cce{} for all datasets.

This performance gap is particularly pronounced on the \textit{Megamarket}, \textit{Zvuk}, and \textit{Beauty} datasets. Notably, on \textit{Beauty}, using \ccem{} or \cce{} leads to improvements in NDCG@10 by 215.5\% and 182.8\%, respectively, and in Coverage@10 by 4933\% and 3106\%.

The Surprisal@10 metric is also lower for all datasets, with the exception of the \textit{Movielens-20m} dataset. For this dataset, the metrics are higher by 8.1\% relative to \ccem{} and by 9.0\% relative to \cce{}.

However, since BCE uses only one negative sample, training with this loss function requires considerably less time per epoch than \ccem{} and \cce{}. Higher slowdown is 1182\% for the \textit{30Music} dataset. For \textit{Megamarket}, a usage of \cce{} leads to a rise in training epoch by 3935.9\% compared with \bce{}. Notably, memory consumption reduces by 8.5\%. Memory consumption for \cce{} is higher only for training on \textit{30Music}, \textit{Gowalla}, \textit{Beauty}. On the other hand, \ccem{} requires more memory than \bce{} for all datasets with the exception \textit{Movielens-20M}.

However, since the \bce{} loss uses only a single negative sample per positive instance, training with this loss is significantly faster per epoch compared to \ccem{} and \cce{}. The highest observed slowdown for \ccem{} is 1182\% on the \textit{30Music} dataset. Conversely, for the \textit{Megamarket} dataset, using \cce{} increases the training time per epoch by 3935.9\% compared to \bce{}. Despite this slowdown, using \cce{} leads to only an 8.5\% decrease in memory efficiency.

In fact, \cce{} requires more memory than \bce{} only on the \textit{30Music}, \textit{Gowalla}, and \textit{Beauty} datasets. In contrast, \ccem{} consistently demands more memory across all datasets except \textit{Movielens-20M}.

\begin{table*}[h]
\centering
\small
\caption{Results for the \textbf{BCE} loss with the SASRec model. }
\label{tab:sasrec_bce}
\renewcommand{\arraystretch}{1.2}
\resizebox{0.85\textwidth}{!}{
\begin{tabular}{lcccccc}
\hline
\textbf{Dataset} & \textbf{Parameters} & \multicolumn{5}{c}{\textbf{BCE}} \\
\cmidrule(lr){3-7}
& \textbf{(bs/sl)} & \textbf{NDCG@10} & \textbf{Coverage@10} & \textbf{Surprisal@10} & \textbf{Mem, GB} & \textbf{Time, s} \\
\hline
Movielens-20m   & 64   / 512 & 0.0300  & 0.1893  & 0.0875  & 1.5    & 66.9   \\
Movielens-20m   & 256  / 512 & 0.0329  & 0.1903  & 0.0871  & 3.5    & 38.3   \\
Beauty          & 1024 / 32  & 0.0037  & 0.0036  & 0.4529  & 2.4    & 30.4   \\
Gowalla         & 512  / 256 & 0.0175  & 0.1218  & 0.5693  & 4.7    & 14.2   \\
Gowalla         & 1024 / 320 & 0.0198  & 0.1273  & 0.5686  & 9.5    & 14.3   \\
Zvuk            & 1024 / 32  & 0.0250  & 0.0552  & 0.2994  & 9.2    & 34.8   \\
Zvuk            & 1024 / 128 & 0.0307  & 0.0782  & 0.3360  & 11.2   & 31.8   \\
30Music         & 256  / 128 & 0.0803  & 0.0587  & 0.4809  & 6.5    & 8.9    \\
30Music         & 256  / 720 & 0.0912  & 0.0679  & 0.4981  & 8.8    & 21.6   \\
Megamarket      & 1024 / 32  & 0.0043  & 0.0492  & 0.3767  & 14.3   & 26.9   \\
Megamarket      & 5120 / 320 & 0.0071  & 0.0559  & 0.3788  & 60.1   & 55.7   \\
\hline
\end{tabular}
}
\end{table*}

\begin{table*}[h]
\centering
\small
\caption{Results for the comparison of \textbf{BCE} loss with \ccem{} for the SASRec model. \(\Delta\): Relative difference compared to \textbf{BCE} and \ccem{}, green denotes \ccem{} superiority.}
\label{tab:sasrec_bce_ccem}
\renewcommand{\arraystretch}{1.2}
\resizebox{0.85\textwidth}{!}{
\begin{tabular}{lcccccc}
\hline
& \textbf{(bs/sl)} &  \textbf{$\Delta$ NDCG, \%} &  \textbf{$\Delta$ Coverage., \%} &  \textbf{$\Delta$ Surprisal., \%} & \textbf{$\Delta$ Mem, \%} & \textbf{$\Delta$ Time, \%} \\
\hline
Movielens-20m   & 256  / 512 &  \textcolor{green!50!black}{↑ 58.9\%}  &  \textcolor{green!50!black}{↑ 27.7\%}  &  \textcolor{red}{↓ 8.1\%}  &    0.0\%  &  \textcolor{red}{↑ 35.8\%}  \\
Beauty          & 1024 / 32  &  \textcolor{green!50!black}{↑ 215.5\%}  &  \textcolor{green!50!black}{↑ 4933.3\%}  &  \textcolor{green!50!black}{↑ 11.9\%}  &  \textcolor{red}{↑ 12.5\%}  &  \textcolor{green!50!black}{↓ 24.3\%} \\
Gowalla         & 1024 / 320 &  \textcolor{green!50!black}{↑ 41.2\%}  &  \textcolor{green!50!black}{↑ 117.6\%}  &  \textcolor{green!50!black}{↑ 10.4\%}  &  \textcolor{red}{↑ 16.8\%}  &  \textcolor{red}{↑ 46.9\%}  \\
Zvuk            & 1024 / 128 &  \textcolor{green!50!black}{↑ 177.8\%}  &  \textcolor{green!50!black}{↑ 90.4\%}  &  \textcolor{green!50!black}{↑ 3.7\%}  &  \textcolor{red}{↑ 188.4\%}  &  \textcolor{red}{↑ 868.6\%}  \\
30Music         & 256  / 720 &  \textcolor{green!50!black}{↑ 57.4\%}  &  \textcolor{green!50!black}{↑ 82.7\%}  &  \textcolor{green!50!black}{↑ 17.1\%}  &  \textcolor{red}{↑ 563.6\%}  &  \textcolor{red}{↑ 1182.4\%}  \\
Megamarket      & 5120 / 320 &  \textcolor{green!50!black}{↑ 135.7\%}  &  \textcolor{green!50!black}{↑ 107.2\%}  &  \textcolor{green!50!black}{↑ 30.4\%}  &  \textcolor{red}{↑ 29.8\%}  &  \textcolor{red}{↑ 370.4\%}  \\
\hline
\end{tabular}
}
\end{table*}

\begin{table*}[h]
\centering
\small
\caption{Results for the comparison of \textbf{BCE} loss with \cce{} for the SASRec model. \(\Delta\): Relative difference compared to \textbf{BCE} and \cce{}, green denotes \cce{} superiority.}
\label{tab:sasrec_bce_cce}
\renewcommand{\arraystretch}{1.2}
\resizebox{0.85\textwidth}{!}{
\begin{tabular}{lcccccc}
\hline
& \textbf{(bs/sl)} &  \textbf{$\Delta$ NDCG, \%} &  \textbf{$\Delta$ Coverage., \%} &  \textbf{$\Delta$ Surprisal., \%} & \textbf{$\Delta$ Mem, \%} & \textbf{$\Delta$ Time, \%} \\
\hline
Movielens-20m   & 256  / 512 &  \textcolor{green!50!black}{↑ 55.3\%}  &  \textcolor{green!50!black}{↑ 10.5\%}  &  \textcolor{red}{↓ 9.0\%}  &  \textcolor{green!50!black}{↓ 2.9\%}  &  \textcolor{red}{↑ 12.3\%}  \\
Beauty          & 1024 / 32  &  \textcolor{green!50!black}{↑ 182.8\%}  &  \textcolor{green!50!black}{↑ 3106.6\%}  &  \textcolor{green!50!black}{↑ 14.2\%}  &  \textcolor{red}{↑ 12.5\%}  &  \textcolor{red}{↑ 164.1\%}  \\
Gowalla         & 1024 / 320 &  \textcolor{green!50!black}{↑ 21.9\%}  &  \textcolor{green!50!black}{↑ 58.8\%}  &  \textcolor{green!50!black}{↑ 13.4\%}  &  \textcolor{red}{↑ 24.2\%}  &  \textcolor{red}{↑ 823.1\%}  \\
Zvuk            & 1024 / 128 &  \textcolor{green!50!black}{↑ 206.5\%}  &  \textcolor{green!50!black}{↑ 79.1\%}  &  \textcolor{green!50!black}{↑ 5.6\%}  &  \textcolor{green!50!black}{↓ 5.4\%}  &  \textcolor{red}{↑ 1629.6\%}  \\
30Music         & 256  / 720 &  \textcolor{green!50!black}{↑ 52.7\%}  &  \textcolor{green!50!black}{↑ 77.1\%}  &  \textcolor{green!50!black}{↑ 16.1\%}  &  \textcolor{red}{↑ 6.8\%}  &  \textcolor{red}{↑ 1700.9\%}  \\
Megamarket      & 5120 / 320 &  \textcolor{green!50!black}{↑ 154.0\%}  &  \textcolor{green!50!black}{↑ 205.1\%}  &  \textcolor{green!50!black}{↑ 44.3\%}  &  \textcolor{green!50!black}{↓ 8.5\%}  &  \textcolor{red}{↑ 3935.9\%}  \\
\hline
\end{tabular}
}
\end{table*}




\subsection{\texorpdfstring{Analysis of efficiency of \cce{} and \ccem{} for Bert4Rec}{Analysis of efficiency of CCE and CCE- for Bert4Rec}}
\label{sec: efficiency_cce_ccem_for_Bert4Rec}

In addition to the SASRec architecture, we study the efficiency of \cce{} and \ccem{} for Bert4Rec ~\cite{sun2019bert4rec}.

Table~\ref{tab:bert4rec_cce_all_metrics} in Appendix~\ref{app:bert4rec_results} presents a comparative analysis of \cce{} and \ccem{} on the Bert4Rec model using an identical batch size and sequence length. It is evident that \ccem{} allows for a reduction in training time on most datasets due to negative sampling. For example, on \textit{Megamarket}, there is a 3.8\% improvement in quality $\mathrm{NDCG@10}$ alongside a 14.4\% decrease in memory consumption and a 32.7\% reduction in training time. The quality gain is especially pronounced on \textit{Gowalla} (68.5\%). At the same time, some datasets (\textit{Zvuk}, \textit{30Music}) show a slight drop in quality metrics with a significant time gain.

For completeness, we also provide metrics for Bert4Rec trained with \ce{}, \cem{} in Tables~\ref{tab:ce_bert4rec}, \ref{tab:cem_bert4rec} in Appendix~\ref{app:bert4rec_results}. Table~\ref{tab:ce_bert4rec} indicates that using the full \ce{} achieves quality comparable to \cce{} on datasets with relatively small item catalogs, such as \textit{Movielens-20m}, \textit{Beauty,} and \textit{Gowalla}. On datasets with large item catalogs, the performance of \ce{} is considerably worse than that of \cce{}, which enables the use of significantly larger batch sizes and sequence lengths.

According to Table~\ref{tab:cem_bert4rec}, \cem{} exhibits nearly zero NDCG@10 values on most datasets, indicating that this loss is unsuitable for the task. In contrast, \ccem{} shows results comparable to \ce. This may be due to batch size and the number of negatives: smaller batch sizes and fewer negatives lead to a strong drop in $\mathrm{NDCG@10}$ quality. 

The results demonstrate that \cce{} and \ccem{} are effective for Bert4Rec. These approaches improve the model's predictive quality on datasets with large catalogs by enabling the use of larger batch sizes and longer sequence lengths through memory optimization.









\section{Conclusions}

\emph{We proposed \cce{} and \ccem{}} for memory-efficient \sr{} training as a replacement for full \ce{} loss and its variant with negative sampling correspondingly. By employing \emph{efficient Triton kernel implementations}, these methods eliminate the memory bottleneck caused by materializing logit tensors. Our experiments show that \emph{\cce{} and \ccem{} reduce the training memory} footprint by 75–97\% and \emph{accelerate training} by 23–71\% if compared to the conventional PyTorch implementation. 
Notably, \ccem{}'s ability to reduce the number of negative samples can further \emph{decrease epoch duration and enhance model performance} compared to \cce{}, and \cce{} can be further optimized through gradient pruning in the final layer, providing a complementary strategy to minimize computational overhead. 

Future research could focus on enhancing the computational efficiency of Triton kernels, integrating alternative negative sampling strategies, and developing adaptive gradient pruning techniques to balance sparsity and model performance. 

Our experiments with SASRec and BERT4Rec demonstrate that memory optimization during training using \ce{}-type losses can enhance sequential recommender systems. The insights and optimizations we employ, such as sparse gradients in  \cce{} and the design of \ccem{}, are model-agnostic and can be seamlessly integrated into other architectures, ensuring the broader applicability of our findings to \sr{}.

\clearpage

\section{Acknowledgements}

The work was supported by the grant for research centers in the field of AI provided by the Ministry of Economic Development of the Russian Federation in accordance with the agreement 000000C313925P4F0002 and the agreement with Skoltech №139-10-2025-033.


\clearpage
\bibliographystyle{ACM-Reference-Format}
\bibliography{main}
\newpage
{\color{gray}\hrule}
\section*{Appendix}
\vspace{1em}
{\color{gray}\hrule}
\vspace{1em}
\appendix
\vspace{1em}
\section{Comparison with Alternative Training Acceleration Techniques}
\label{app:alternatives}

Training acceleration for sequential recommendation models can be approached from several angles:
\emph{precision reduction} (FP16/BF16 mixed precision, INT8 quantization), \emph{weight sparsification}
(2:4 semi-structured pruning~\cite{hu2024accelerating}), and \emph{gradient-level methods} such as the
sparse gradients and fused kernels proposed in this work.
A key distinction often overlooked in practice is whether a technique targets \emph{inference} or
\emph{training}, and whether it reduces \emph{memory} or \emph{wall-clock training time}.
For example, gradient checkpointing and embedding quantization primarily reduce memory footprint but do
not accelerate training throughput; several quantization schemes are designed for inference only and
offer no benefit during the backward pass~\cite{wortsman2023stable}.

\subsection*{The Gradient Computation Bottleneck}

Any technique that accelerates only the forward pass captures a fundamentally limited share of total
training time, since gradient computation dominates.
On MovieLens-20M, the forward pass accounts for approximately 29\% of total training time for SASRec
and 35\% for BERT4Rec, with the backward pass consuming the remaining 71\% and 65\%, respectively.
This asymmetry means that even a $2\times$ speedup on the forward pass alone would reduce total
training time by at most 15--18\%.
Methods that target the backward pass — or eliminate redundant computation in both passes jointly,
as CCE and CCE\textsuperscript{-} do — are therefore considerably more impactful.

\subsection*{Quantization and Mixed Precision: Limitations for Training}

Table~\ref{tab:matmul_timing} reports wall-clock times for the classification head matrix multiplication
under different precision regimes on the Megamarket and Zvuk datasets, isolating the operation that
dominates memory and compute at the final layer.
W8A8-INT8 quantization achieves a ${\sim}2.6\times$ speedup over dense FP16 on this operation.
However, two important caveats apply.
First, this speedup is hardware-dependent: acceleration was observed on A40 GPUs but not on A100s,
due to the small hidden dimension ($D{=}320$) and differences in tensor core utilisation between
architectures.
Second, and more fundamentally, mixed-precision and quantization schemes still \emph{materialise the
full logit tensor} of size $bs \times sl \times |V|$ during the forward pass, and the backward pass
continues to accumulate gradients in FP32.
As a result, neither approach resolves the memory bottleneck that makes training with CE loss on
large catalogs prohibitive.

Table~\ref{tab:method_comparison} summarises all considered approaches.
\textbf{(1) FP16/BF16 mixed precision offers a reliable ${\sim}2\times$ speedup over FP32 with negligible accuracy
loss} and is compatible with negative sampling; it is already assumed as the baseline in all our main
experiments.
\textbf{(2) INT8 quantization provides additional speed on compatible hardware but incurs accuracy drops}
exceeding 30\% on Megamarket, making it unsuitable as a general-purpose training strategy for
large-catalog datasets.
\textbf{(3) Semi-structured 2:4 sparsification yields a ${\sim}1.6\times$ speedup and preserves accuracy}.
CCE and CCE\textsuperscript{-} are unique in simultaneously addressing memory (75--97\% reduction)
and training speed (23--89\% reduction) while remaining fully compatible with the negative sampling
strategies that are essential for large catalogs.

\begin{table}[t]
\centering
\caption{\textbf{Bert4Rec - Classification Head Acceleration.} Wall-clock time for the classification head matrix multiplication under different precision
and sparsity regimes. Speedup is relative to dense FP16. INT8 acceleration is observed on A40 GPUs,
no speedup was measured on A100s due to hardware-specific tensor core constraints.}
\label{tab:matmul_timing}
\begin{tabular}{llcccc}
\toprule
\textbf{Method} & \textbf{Dataset} & \textbf{Input size} & \textbf{Weight size} & \textbf{Time (s)} & \textbf{Speedup} \\
\midrule
Dense FP16          & Megamarket & $(1600, 320)$  & $(1{,}609{,}625,\ 320)$ & 41.22 & $1.00\times$ \\
Semi-struct. 2:4    & Megamarket & $(1600, 320)$  & $(1{,}609{,}632,\ 256)$ & 25.84 & $1.56\times$ \\
Quant. W8A8-INT8    & Megamarket & $(1600, 320)$  & $(1{,}609{,}632,\ 256)$ & 15.67 & $2.61\times$ \\
\midrule
Dense FP16          & Megamarket & $(100,\ 320)$  & $(1{,}609{,}625,\ 256)$ & 3.37  & $1.00\times$ \\
Semi-struct. 2:4    & Megamarket & $(100,\ 320)$  & $(1{,}609{,}632,\ 256)$ & 1.91  & $1.76\times$ \\
\midrule
Dense FP16          & Zvuk       & $(3200, 320)$  & $(887{,}045,\ 320)$     & 43.80 & $1.00\times$ \\
Semi-struct. 2:4    & Zvuk       & $(3200, 320)$  & $(887{,}072,\ 320)$     & 22.95 & $1.91\times$ \\
Quant. W8A8-INT8    & Zvuk       & $(3200, 320)$  & $(887{,}072,\ 320)$     & 16.53 & $2.65\times$ \\
\midrule
Dense FP16          & Zvuk       & $(100,\ 320)$  & $(887{,}045,\ 320)$     & 1.94  & $1.00\times$ \\
Semi-struct. 2:4    & Zvuk       & $(100,\ 320)$  & $(887{,}072,\ 320)$     & 1.06  & $1.83\times$ \\
\bottomrule
\end{tabular}
\end{table}

\begin{table}[t]
\centering
\caption{Qualitative comparison of training acceleration methods for Bert4Rec and SASRec (results are averaged).
Speedups are reported relative to dense FP32 training unless otherwise noted. Accuracy drop is
assessed on large-catalog datasets (Megamarket, Zvuk). $\checkmark$ / $\times$ indicates whether
the method is directly compatible with negative sampling during training.}
\label{tab:method_comparison}
\small
\begin{tabular}{lccccl}
\toprule
\textbf{Method} & \textbf{Speedup} & \textbf{Memory reduction} & \textbf{Acc.\ drop} & \textbf{Neg.\ sampling} & \textbf{Notes} \\
\midrule
FP16/BF16 mixed precision   & ${\sim}2\times$   & minimal   & ${<}1\%$    & $\checkmark$ & Baseline in all our experiments \\
INT8 quantization (W8A8)    & ${\sim}2.6\times$ & minimal   & ${>}30\%$\textsuperscript{$\dagger$}  & $\checkmark$ & A40 only; no gain on A100 \\
Semi-struct.\ 2:4 sparsity  & ${\sim}1.6\times$ & minimal   & ${<}1\%$    & $\times$     & Forward pass only \\
Sparse gradients (CCE)      & ${\sim}1.6\times$ & ${\sim}90\%$ & ${<}1\%$ & $\times$     & Gradient filtering at final layer \\
CCE                         & ${\sim}1.6\times$ & $75$--$97\%$ & ${<}1\%$ & $\times$     & Fused kernel; no neg.\ sampling \\
CCE\textsuperscript{-} (ours) & up to ${\sim}2\times$ & $75$--$97\%$ & ${<}1\%$ & $\checkmark$ & Fused kernel with neg.\ sampling \\
\bottomrule
\end{tabular}
\smallskip

\noindent\textsuperscript{$\dagger$}Accuracy drop ${<}3\%$ observed on Zvuk; ${>}30\%$ on Megamarket.
\end{table}

\section{Practical Scaling Laws}
\label{app:scalinglaws}
Fixing $sl$ and treating the memory budget as $M = bs \cdot ns$, the 
Lagrangian optimality condition yields:
\[
    \alpha \cdot ns^{NS} = \beta \cdot bs^{BS}, 
    \quad \alpha = C \cdot NS,\ \beta = D \cdot BS,
\]
which, combined with the constraint $bs \cdot ns = M$, gives:
\[
    \frac{ns^*}{bs^*} = 
    \left(\frac{\beta}{\alpha}\right)^{\!\frac{2}{NS+BS}} 
    \cdot M^{\,\gamma}, 
    \qquad \gamma = \frac{BS - NS}{NS + BS}.
\]
The exponent $\gamma$ determines how the optimal ratio scales with the 
memory budget. When $\gamma = 0$ (i.e., $BS = NS$), the ratio is 
independent of $M$ and equals $\alpha/\beta$. Table~\ref{tab:lagrange} 
reports the fitted quantities for all datasets. MovieLens-20M is the 
only dataset where $\gamma = 0$, yielding a constant optimal ratio of 
$ns^*/bs^* \approx 1.42$, meaning negative samples should slightly 
outnumber batch elements. For all other datasets $\gamma \neq 0$, so 
the optimal split depends on the available memory budget and should be 
computed per-dataset using the fitted coefficients. Megamarket 
($\gamma = -11.67$) is an exception: the near-zero denominator 
$NS + BS \approx -0.02$ makes this estimate numerically unstable and 
should not be used for practical allocation.

\begin{table}[ht]
\centering
\caption{Lagrangian quantities for the optimal $ns$/$bs$ allocation 
under a fixed memory budget $M = bs \cdot ns$. $\gamma = (BS-NS)/(NS+BS)$ 
controls how the optimal ratio scales with $M$; $\gamma=0$ implies a 
constant ratio independent of $M$.}
\label{tab:lagrange}
\begin{tabular}{lrrrrl}
\toprule
Dataset & $\alpha = C \cdot NS$ & $\beta = D \cdot BS$ & $NS+BS$ & $\gamma$ & Note \\
\midrule
Zvuk          &  0.055 &  1.055 & $-$0.472 & $+$1.372 & ratio grows with $M$ \\
Megamarket    &  0.163 &  0.107 & $-$0.019 & $-$11.67 & unstable (near-zero denom.) \\
30Music       &  0.332 &  0.429 & $-$0.480 & $-$0.036 & ratio $\approx$ constant \\
Beauty        &  0.994 &  0.124 & $-$0.362 & $-$1.459 & ratio shrinks with $M$ \\
Gowalla       &  0.556 &  1.434 & $-$1.225 & $-$0.115 & ratio $\approx$ constant \\
Movielens-20m &  3.950 &  2.783 & $-$2.000 & $\phantom{+}$0.000 & $ns^*/bs^* = 1.42$ (constant) \\
\bottomrule
\end{tabular}
\end{table}

\section{Bert4Rec Results}
\label{app:bert4rec_results}

\begin{table*}[h]
\centering
\small
\caption{Comparison of NDCG@10, memory consumption, and epoch time for \cce{} and \ccem{} on the Bert4Rec model using identical training hyperparameters. Bold values indicate the best performance per dataset. \(\Delta\): Relative difference compared to \cce{}, green denotes \ccem{} superiority.}
\label{tab:bert4rec_cce_all_metrics}
\renewcommand{\arraystretch}{1.2}
\resizebox{0.85\textwidth}{!}{
\begin{tabular}{llccccccccc}
\hline
\textbf{Dataset} & \textbf{Parameters}
& \multicolumn{3}{c}{\textbf{NDCG@10}}
& \multicolumn{3}{c}{\textbf{Mem, GB}}
& \multicolumn{3}{c}{\textbf{Time, s}} \\
\cmidrule(lr){3-5} \cmidrule(lr){6-8} \cmidrule(lr){9-11}
& \textbf{(bs/sl/ns)}
& \cce{} & \ccem{} & \textbf{$\Delta$, \%}
& \cce{} & \ccem{} & \textbf{$\Delta$, \%}
& \cce{} & \ccem{} & \textbf{$\Delta$, \%} \\
\hline
Movielens-20m & 256 / 512 / 511
    & \textbf{0.0489} & 0.0481 & \textcolor{red}{↓ 1.5\%}
    & \textbf{3.5} & \textbf{3.5} & 0\%
    & \textbf{24.3} & 26.3 & \textcolor{red}{↑ 8.2\%} \\

Beauty & 1024 / 32 / 511
    & 0.0080 & \textbf{0.0085} & \textcolor{green!50!black}{↑ 6.3\%}
    & \textbf{2.1} & \textbf{2.1} & 0\%
    & \textbf{21.0} & 21.3 & \textcolor{red}{↑ 1.4\%} \\

Gowalla & 1024 / 320 / 511
    & 0.0092 & \textbf{0.0155} & \textcolor{green!50!black}{↑ 68.5\%}
    & \textbf{10.7} & \textbf{10.7} & 0\%
    & 13.0 & \textbf{10.6} & \textcolor{green!50!black}{↓ 18.5\%} \\

Zvuk & 1024 / 128 / 2047
    & \textbf{0.0793} & 0.0639 & \textcolor{red}{↓ 19.5\%}
    & \textbf{10.2} & 10.6 & \textcolor{red}{↑ 3.9\%}
    & 101.6 & \textbf{65.8} & \textcolor{green!50!black}{↓ 35.2\%} \\

30Music & 256 / 720 / 4095
    & \textbf{0.0885} & 0.0803 & \textcolor{red}{↓ 9.2\%}
    & \textbf{11.6} & 12.3 & \textcolor{red}{↑ 6.0\%}
    & 55.9 & \textbf{51.1} & \textcolor{green!50!black}{↓ 8.6\%} \\

Megamarket & 2048 / 320 / 4095
    & 0.0125 & \textbf{0.0130} & \textcolor{green!50!black}{↑ 3.8\%}
    & 62.8 & \textbf{53.8} & \textcolor{green!50!black}{↓ 14.4\%}
    & 306.3 & \textbf{206.1} & \textcolor{green!50!black}{↓ 32.7\%} \\
\hline
\end{tabular}
}
\end{table*}

\begin{table*}[th]
\centering
\small
\caption{Comparison of Coverage@10 and Surprisal@10 for \cce{} and \ccem{} on the Bert4Rec model using identical hyperparameters. Bold values indicate the best performance per dataset. \(\Delta\): Relative difference compared to \cce{}, green denotes \ccem{} superiority.}
\label{tab:new_cce_cov_surp}
\renewcommand{\arraystretch}{1.2}
\resizebox{0.85\textwidth}{!}{
\begin{tabular}{llcccccc}
\hline
\textbf{Dataset} & \textbf{Parameters}
& \multicolumn{3}{c}{\textbf{Coverage@10}}
& \multicolumn{3}{c}{\textbf{Surprisal@10}} \\
\cmidrule(lr){3-5} \cmidrule(lr){6-8}
& \textbf{(bs/sl/ns)}
& \cce{} & \ccem{} & \textbf{$\Delta$, \%}
& \cce{} & \ccem{} & \textbf{$\Delta$, \%} \\
\hline
Movielens-20m & 256 / 512 / 511
    & \textbf{0.1690} & 0.1617 & \textcolor{red}{↓ 4.3\%}
    & \textbf{0.0744} & 0.0724 & \textcolor{red}{↓ 2.8\%} \\

Beauty & 1024 / 32 / 511
    & \textbf{0.0502} & 0.0301 & \textcolor{red}{↓ 40.0\%}
    & \textbf{0.5093} & 0.4796 & \textcolor{red}{↓ 5.8\%} \\

Gowalla & 1024 / 320 / 511
    & \textbf{0.1710} & 0.1557 & \textcolor{red}{↓ 9.0\%}
    & \textbf{0.6548} & 0.6109 & \textcolor{red}{↓ 6.7\%} \\

Zvuk & 1024 / 128 / 2047
    & \textbf{0.1042} & 0.1014 & \textcolor{red}{↓ 2.7\%}
    & \textbf{0.3578} & 0.3435 & \textcolor{red}{↓ 4.0\%} \\

30Music & 256 / 720 / 4095
    & 0.0799 & \textbf{0.0849} & \textcolor{green!50!black}{↑ 6.3\%}
    & \textbf{0.5726} & 0.5600 & \textcolor{red}{↓ 2.2\%} \\

Megamarket & 2048 / 320 / 4095
    & \textbf{0.1410} & 0.1352 & \textcolor{red}{↓ 4.1\%}
    & \textbf{0.5603} & 0.5392 & \textcolor{red}{↓ 3.8\%} \\
\hline
\end{tabular}
}
\end{table*}

\begin{table*}[h]
\centering
\small
\caption{Performance of Bert4Rec with \ce{} loss.}
\label{tab:ce_bert4rec}
\renewcommand{\arraystretch}{1.2}
\resizebox{0.85\textwidth}{!}{
\begin{tabular}{llccccc}
\hline
\textbf{Dataset} & \textbf{Parameters}
& \textbf{NDCG@10}
& \textbf{Coverage@10}
& \textbf{Surprisal@10}
& \textbf{Mem, GB}
& \textbf{Time, s} \\
\hline
Movielens-20m & 256 / 512 & 0.0487 & 0.1890 & 0.0771 & 7.0 & 32.9 \\
Beauty & 1024 / 32 & 0.0083 & 0.0316 & 0.4875 & 12.6 & 62.8 \\
Gowalla & 256 / 64 & 0.0091 & 0.0953 & 0.6321 & 12.9 & 30.0 \\
Zvuk & 64 / 96 & 0.0507 & 0.0221 & 0.2644 & 38.1 & 1514.6 \\
30Music & 32 / 128 & 0.0237 & 0.0087 & 0.4026 & 15.4 & 183.2 \\
Megamarket & 32 / 128 & 0.0035 & 0.0044 & 0.3603 & 62.7 & 3684.4 \\
\hline
\end{tabular}
}
\end{table*}

\begin{table*}[h]
\centering
\small
\caption{Performance of Bert4Rec with \cem{} loss.}
\label{tab:cem_bert4rec}
\renewcommand{\arraystretch}{1.2}
\resizebox{0.85\textwidth}{!}{
\begin{tabular}{llccccc}
\hline
\textbf{Dataset} & \textbf{Parameters}
& \textbf{NDCG@10}
& \textbf{Coverage@10}
& \textbf{Surprisal@10}
& \textbf{Mem, GB}
& \textbf{Time, s} \\
\hline
Movielens-20m & 64 / 512 / 1512 & 0.0002 & 0.0257 & 0.1928 & 1.0 & 35.6 \\
Beauty & 1024 / 32 / 512 & 0.0000 & 0.0009 & 0.7653 & 2.8 & 21.2 \\
Gowalla & 512 / 256 / 512 & 0.0001 & 0.0009 & 0.6236 & 9.9 & 13.2 \\
Zvuk & 1024 / 32 / 256 & 0.0001 & 0.0006 & 0.4146 & 22.8 & 86.4 \\
30Music & 256 / 128 / 511 & 0.0000 & 0.0000 & 0.6132 & 23.4 & 40.5 \\
Megamarket & 1024 / 32 / 255 & 0.0040 & 0.0415 & 0.4278 & 35.4 & 67.1 \\
\hline
\end{tabular}
}
\end{table*}

\section{Additional results}
\label{app:additional_results}

\begin{figure*}[h]
  \centering
  \includegraphics[width=0.98\textwidth]{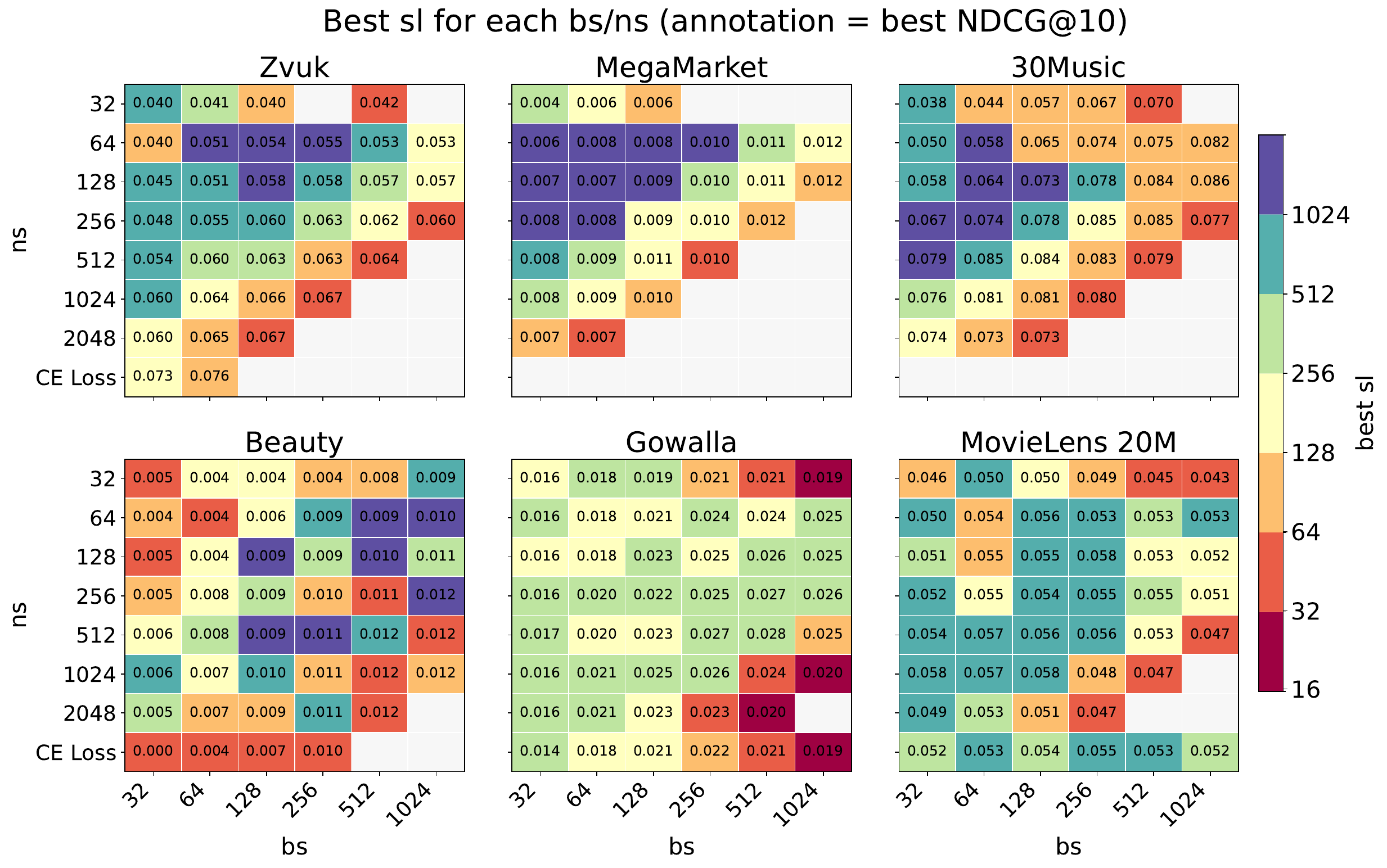}
  \caption{Best sequence length ($sl$) for each combination of batch size ($bs$) and the number of negative samples ($ns$). Color encodes the winning $sl$, while each cell annotation reports the best achieved \textbf{NDCG} for that $(bs, ns)$ configuration. The `CE Loss' label corresponds to runs using the full cross-entropy loss instead of sampled negatives.}
  \label{fig:best_sl_map_ndcg}
  \Description{description}
\end{figure*}

\begin{figure*}[h]
  \centering
  \includegraphics[width=0.98\textwidth]{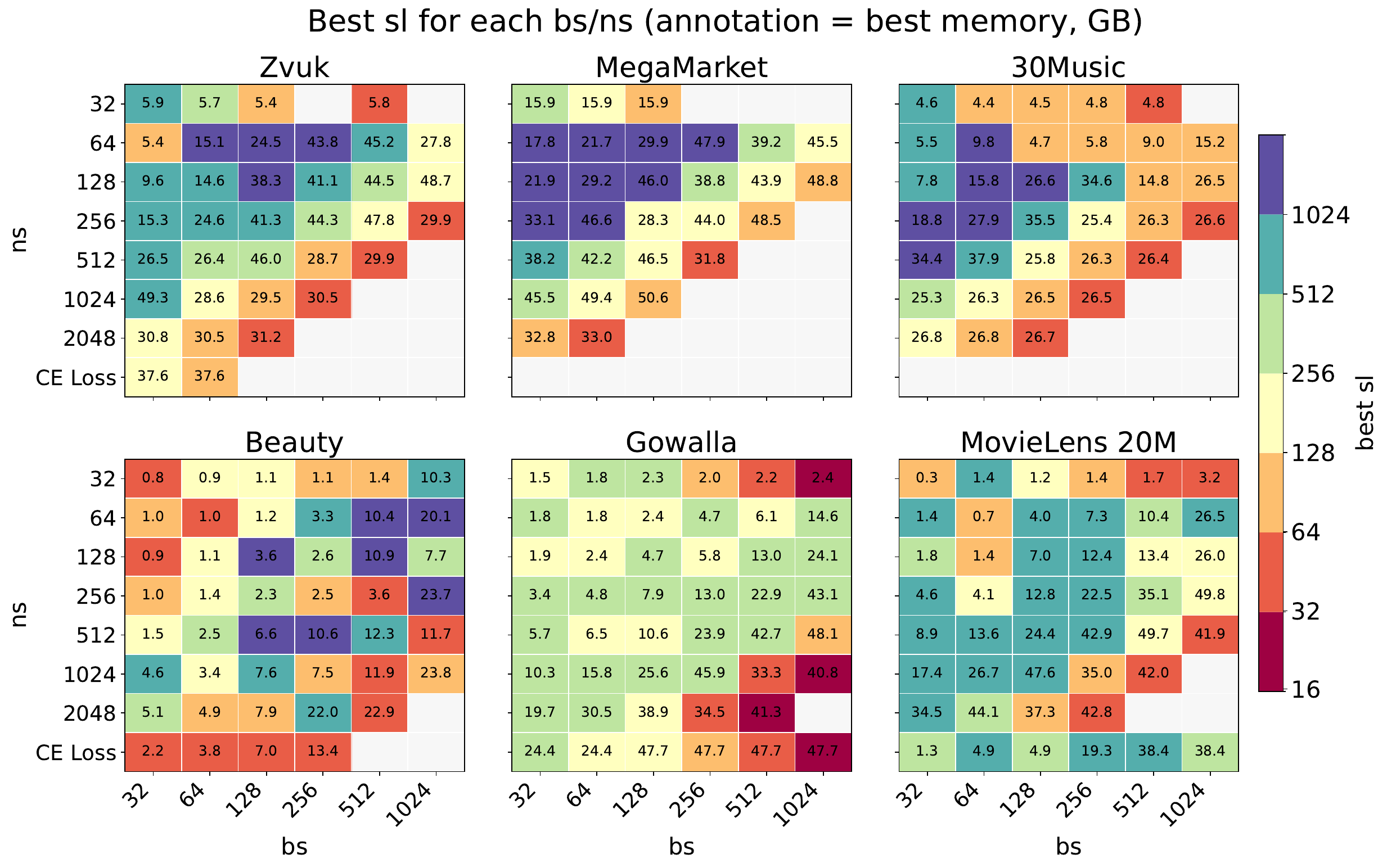}
  \caption{Best sequence length ($sl$) for each combination of batch size ($bs$) and the number of negative samples ($ns$). Color encodes the winning $sl$, while each cell annotation reports the memory consumption of the best configuration for that $(bs, ns)$ pair. The `CE Loss' label corresponds to runs using the full cross-entropy loss instead of sampled negatives.}
  \label{fig:best_sl_map_memory}
  \Description{description}
\end{figure*}

Figures~\ref{fig:best_sl_map_ndcg} and~\ref{fig:best_sl_map_memory} summarize which sequence length performs best for each combination of batch size and the number of negative samples. The two views differ only in the cell annotations: Figure~\ref{fig:best_sl_map_ndcg} reports the best achieved \textbf{NDCG}, while Figure~\ref{fig:best_sl_map_memory} reports the corresponding memory consumption.

\begin{figure*}[h]
  \centering
  \includegraphics[width=0.98\textwidth]{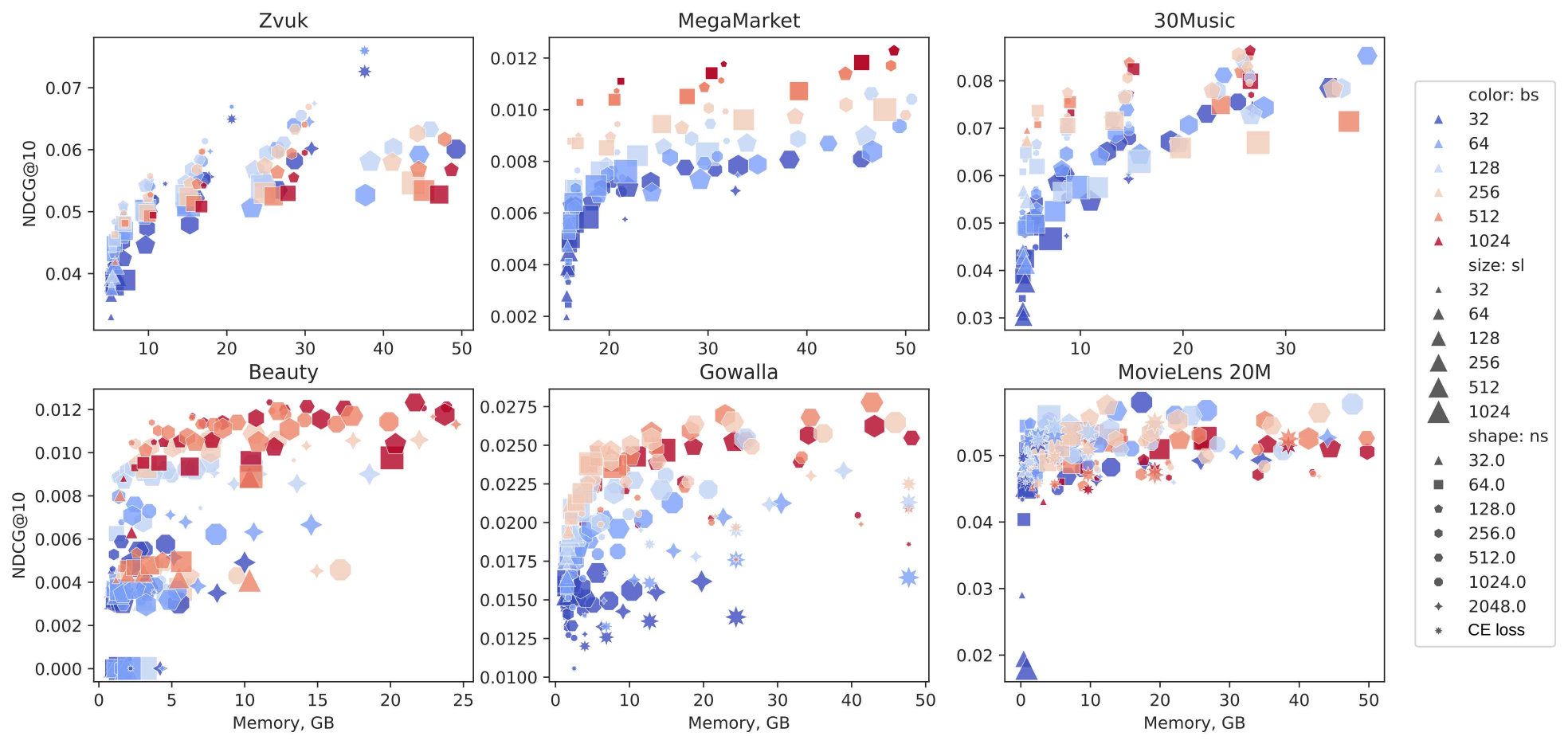}
  \caption{Memory scaling across three dimensions: (1) marker size ($sl$: sequence length), (2) color ($bs$: batch size), and (3) marker shape ($ns$: number of negative samples). Asterisks denote points using the \ce{} loss. For example, brown markers (larger batch sizes) generally yield better average performance, though the largest batch size does not consistently dominate. Higher \textbf{NDCG} occurs with 128--512 negative samples (squares-hexagons); only the Zvuk dataset shows improved performance with asterisks (full \cem{} loss).}
  \label{fig:negative_memory}
  \Description{description}
\end{figure*}

\subsection{Sparsity Threshold Analysis}

\begin{figure}[ht]
  \centering
  \includegraphics[width=0.95\textwidth]{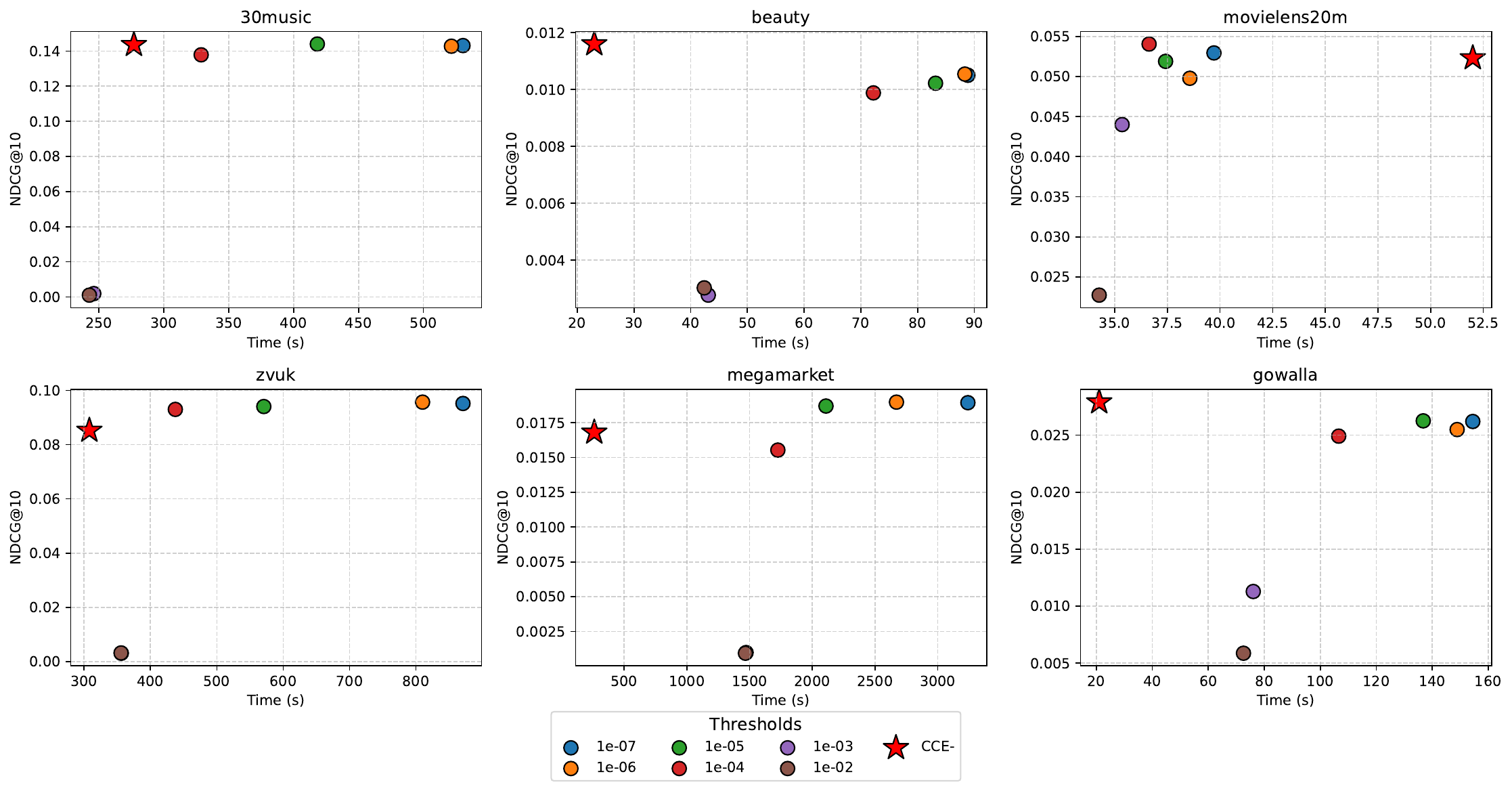}
\caption{NDCG@10 metric and time per epoch of the SASRec models trained with \ccem{} and \cce{} using gradient filtering at increasing thresholds.}
\label{fig:sparse_grads_cce_vs_ccem}
\Description{description}
\end{figure}

\end{document}